\documentclass[10pt,reqno]{article}
\usepackage{graphicx}
\usepackage{indentfirst,csquotes}

\usepackage{amssymb,amsthm,amsmath,bm}
\usepackage{xcolor,hyperref,titlesec}
\usepackage{placeins}  % to add \FloatBarrier
\usepackage{natbib}
\usepackage{titlesec}

\hypersetup{ colorlinks=true, citecolor=black, linkcolor=black, filecolor=black, urlcolor=blue }

\begin{document}
\title{Coating thickness prediction for a viscous film on a rough plate} %%%%%%%%%%%%
\author{
Lebo Molefe\textsuperscript{1,2}, Giuseppe A. Zampogna\textsuperscript{2,}\thanks{giuseppe.zampogna@epfl.ch}, John M. Kolinski\textsuperscript{1}, \and François Gallaire\textsuperscript{2} \\
\small{\textsuperscript{1} Engineering Mechanics of Soft Interfaces Laboratory, EPFL, CH-1015 Lausanne, Switzerland} \\ 
\small{\textsuperscript{2} Laboratory of Fluid Mechanics and Instabilities, EPFL, CH-1015 Lausanne, Switzerland}
}
\date{May 28, 2024}
\maketitle

%%%%%%%%%%

\begin{abstract}
Surface roughness significantly modifies the liquid film thickness entrained when dip coating a solid surface, particularly at low coating velocity. Using a homogenization approach, we present a predictive model for determining the liquid film thickness coated on a rough plate. A homogenized boundary condition at an equivalent flat surface is used to model the rough boundary, accounting for both flow through the rough texture layer, through an interface permeability term, and slip at the equivalent surface. While the slip term accounts for tangential velocity induced by viscous shear stress, accurately predicting the film thickness requires the interface permeability term to account for additional tangential flow driven by pressure gradients along the interface. We find that a greater degree of slip and interface permeability signifies less viscous stress that would promote deposition, thus reducing the amount of free film coated above the textures. The model is found to be in good agreement with experimental measurements and requires no fitting parameters. Furthermore, our model may be applied to arbitrary periodic roughness patterns, opening the door to flexible characterization of surfaces found in natural and industrial coating processes. 
\end{abstract} %%%%%%%%%

\bigskip

\section{Introduction}\label{sec:introduction}

\begin{figure}
    \centering
    \includegraphics[scale=0.19]{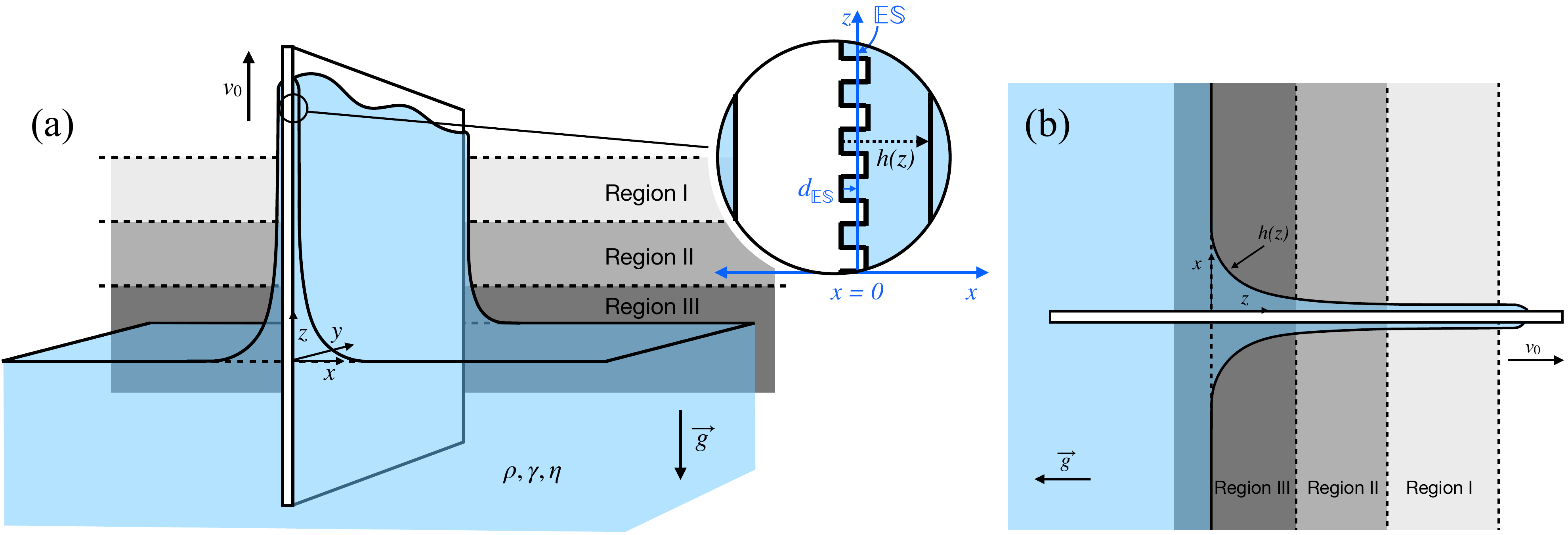}
    \caption{Dip coating system. (a) A solid plate is pulled at constant velocity $\mathbf{v} = v_0 \hat{\mathbf{z}}$ from a liquid bath having density $\rho$, surface tension $\gamma$, and dynamic viscosity $\eta$, where $\mathbf{v}$ is directed opposite the direction of gravitational acceleration $\mathbf{g}$. The plate is rough with a periodic texture pattern. To model the rough surface, we apply an equivalent boundary condition to a flat equivalent surface $\mathbb{ES}$ that is placed a distance $d_{\mathbb{ES}}$ from the bottom of the roughness. The thin film on the plate forms three regions: a flat film (Region I) is connected to a static meniscus (Region III) by a dynamic meniscus (Region II). We wish to know the limiting film thickness $h(z) = h_0$ in the flat film region. (b) The film thickness $h(z)$ as a function of the coordinate $z$ along the plate.}
    \label{fig:dip_coating_system}
\end{figure}

Describing coating flow over a solid is a theoretical challenge, yet endlessly appealing because of the many simple phenomena that involve these kinds of flows: we withdraw spoons covered in honey from a jar, coat strawberries in chocolate, dip and remove a paintbrush from a bucket, lift ourselves out of the water at the swimming pool, and take pictures with film made of a light-sensitive chemical coated over a transparent strip. In general, in dip coating processes, the film thickness $h_0$ coated on objects pulled from a bath of viscous liquid depends on the capillary number $Ca = \frac{\eta v_0}{\gamma}$, which describes the balance between viscous and capillary forces, where $v_0$ is the pulling velocity, $\eta$ is the liquid's dynamic viscosity, and $\gamma$ is the liquid's surface tension. The initial model for dip coating a Newtonian liquid over a smooth, flat plate at small $Ca$ gives the relation \citep{dragging_liquid_moving_plate_1942, on_thickness_layer_liquid_remaining_walls_vessels_after_their_emptying_1943, drag_out_problem_film_coating_theory_1982}
\begin{equation}
    h_0 = 0.94 a Ca^{\frac{2}{3}},
    \label{eq:smooth_surface_film_thickness}
\end{equation}
where $a = \sqrt{\gamma/(\rho g)}$ is the capillary length, $\rho$ is the liquid's density, and $g$ is the magnitude of gravitational acceleration. The model matched earlier experiments of \citet{thickness_liquid_film_adhering_surface_slowly_withdrawn_liquid_1940} who found $h_0 \sim Ca^{\frac{2}{3}}$.

The model \eqref{eq:smooth_surface_film_thickness} has been extended to include corrections accounting for gravity and inertia at larger $Ca$ \citep{free_coating_newtonian_liquid_vertical_surface_1973, gravity_inertia_effects_plate_coating_1998}. Theoretical and experimental extensions have also been made with regard to the object coated -- including tilted plates \citep{drag_out_problem_film_coating_theory_1982, on_drag_out_problem_liquid_film_theory_2008}, horizontal and vertical cylinders \citep{free_coating_newtonian_liquid_vertical_surface_1973, drag_out_problem_film_coating_theory_1982, theory_withdrawal_cylinders_liquid_baths_1966, fluid_coating_fiber_1999}, compliant surfaces \citep{liquid_contact_line_receding_soft_gel_surface_2014, enhanced_dip_coating_soft_substrate_2022}, and rough surfaces \citep{experimental_study_substrate_roughness_surfactant_effects_Landau_Levich_law_2005, coating_textured_solid_2011, liquid_film_entrainment_during_dip_coating_saturated_porous_substrate_2020} -- as well as the liquid properties, as reviewed in \citet{withdrawing_solid_from_bath_how_much_liquid_coated_2017}. Given the multitude of realistic situations in which a surface to be coated is rough or porous, rather than smooth \citep{rhomboidal_lattice_structure_common_feature_sandy_beaches_1976, purity_sacred_lotus_escape_contamination_biological_surfaces_1997, characterization_distribution_water_repellent_self_cleaning_plant_surface_1997, forced_dewetting_porous_media_2007, thin_film_evolution_thin_porous_layer_modeling_tear_film_contact_lens_2010}, we focus our attention on thin film flow over a rough surface pulled from a liquid bath (figure~\ref{fig:dip_coating_system}). The film profile is $h = h(z)$ and in the dynamic meniscus region (Region II), the profile tends toward a flat film of constant thickness $h \to h_0$ as $z \to \infty$. Although initial progress has been made to characterize the effect of roughness on the film thickness \citep{experimental_study_substrate_roughness_surfactant_effects_Landau_Levich_law_2005, entrainments_visqueux_2010, coating_textured_solid_2011, liquid_film_entrainment_during_dip_coating_saturated_porous_substrate_2020}, at present we lack a general model capable of predicting film thickness coated on an arbitrary periodic roughness pattern.

At large $Ca$, the film thickness coated on a rough plate is found to be modeled well by \eqref{eq:smooth_surface_film_thickness}, while experiments demonstrate that at small $Ca$ surface roughness significantly increases film thickness compared to the case of a smooth plate \citep{experimental_study_substrate_roughness_surfactant_effects_Landau_Levich_law_2005, coating_textured_solid_2011}. The physical reason for this increase is that the roughness introduces a minimum film thickness due to viscous resistance to liquid flow against the additional solid walls or pore-like features, which have a greater area compared to the flat case \citep{coating_textured_solid_2011}. For a rough plate, the film thickness no longer goes to zero as $Ca \to 0$ as predicted by \eqref{eq:smooth_surface_film_thickness}; instead, experiments demonstrate that a small film remains trapped within the rough features due to viscous resistance, such that $h_0 \to h_{min}$ as $Ca \to 0$. The scale of this minimum thickness is $h_{min} \sim r$ with $r$ being the typical peak-to-valley scale of the roughness. \citet{coating_textured_solid_2011} thus divide the film into two regions: a trapped fluid region located within the roughness features and a free film region outside, separated by a flat plane located at the peak of the roughness features.

Motivated by such physical reasoning, both \citet{experimental_study_substrate_roughness_surfactant_effects_Landau_Levich_law_2005} and \citet{liquid_film_entrainment_during_dip_coating_saturated_porous_substrate_2020} proposed a model for rough dip coating that would replace the no-slip boundary condition at the rough wall by a slip boundary condition at a fictitious flat boundary plane at the peak of the roughness features. Such a model is sensible if we consider that at the boundary plane we find a mix of solid and fluid patches, meaning that at the boundary there is a nonzero slip velocity \citep{boundary_conditions_naturally_permeable_wall_1967}. Introducing slip has been successfully used to model experimentally observed velocity profiles in channels with rough walls \citep{effective_slip_pressure_driven_stokes_flow_2003, measurement_slip_length_superhydrophobic_surfaces_2012}, where flow velocities are similar to those in small-$Ca$ dip coating experiments. However, \citet{coating_textured_solid_2011} found that a slip model was insufficient to explain their observation of a constant minimum film thickness ($h_0 = h_{min}$) with no free film coated below a critical capillary number $Ca < Ca_c$; the slip model overestimated the  experimentally observed free film thickness. A modified model was developed that proposed to augment the slip velocity by considering porous flow through the rough layer, but it lacked experimental comparison \citep{forced_dewetting_porous_media_2007}. Instead of using a slip model, \citet{coating_textured_solid_2011} proposed to model the system as two layers of liquid, where the trapped-fluid region is replaced by a liquid of higher viscosity $\eta^* > \eta$, and subsequently solve for $h_0$. When $\eta^*$ was used as a fitting parameter to the experimental data, this `two-layer model' predicted the formation of a trapped film and the total depletion of the free film for $Ca < Ca_c$.

Despite the successes of prior modeling efforts \citep{experimental_study_substrate_roughness_surfactant_effects_Landau_Levich_law_2005, coating_textured_solid_2011, liquid_film_entrainment_during_dip_coating_saturated_porous_substrate_2020}, these models are not fully predictive since closure of the problem requires experimental data to fit either the slip length $\mathcal{L}$ or the viscosity increase parameter $\eta^*$. Here, we develop a model that provides further insight into the physical mechanism by which the microstructure affects the macroscopic flow, and, in doing so, present a predictive model for the film thickness coated on rough surfaces by a viscous liquid. The key factor enabling the predictability of the model employed in the present work stems from the use of a homogenization technique \citep{hornung, permeability_rigid_porous_media_2010}, which allows us to derive effective macroscopic properties of a surface from its microscopic structure. The upscaling procedure involves taking a spatial average of microscopic flow quantities describing the flow around a single microscopic roughness feature to compute effective macroscopic properties. Properties upscaled from the microstructure can then be applied to the macroscopic problem via an interface condition imposed over a fictitious flat surface called the `equivalent surface' ($\mathbb{ES}$ in figure~\ref{fig:dip_coating_system}) placed between the trapped liquid layer and the free film.

Homogenization can be used to analyze fluid-solid interaction phenomena where there is a separation of scales between the typical microscopic roughness size and the whole macroscopic system size \citep{permeability_rigid_porous_media_2010}. The large scale may be given by the size of a macroscopic object (such as the radius of a sphere with a rough surface) or by the scale of the flow domain (such as the height of a channel) \citep{generalized_slip_condition_over_rough_surfaces_2019, fluid_flow_over_through_regular_bundle_rigid_fibres_2016}. Thus, homogenization has been extensively employed to calculate the effective flow of an incompressible fluid in domains with no free interface, such as through porous and poroelastic media \citep{carraro,fluid_flow_over_through_regular_bundle_rigid_fibres_2016,lacis,lacis2017computational}, over rough surfaces \citep{jimbol2017,generalized_slip_condition_over_rough_surfaces_2019, transfer_mass_momentum_rough_porous_surfaces_2020, interfacial_conditions_free_fluid_region_porous_medium_2021}, and across periodic and weakly periodic microstructured permeable surfaces \citep{effective_stress_jump_across_membranes_2020,homogenization_based_design_microstructured_membranes_2021,zampogna2022transport}. \citet{botNaq2020} have shown that the flow at an interface between a free-fluid region and a rough surface is described by the slip tensor $\bm{\mathcal{L}}$ and interface permeability tensor $\bm{\mathcal{K}}^{itf}$ \citep[see also][]{interfacial_conditions_free_fluid_region_porous_medium_2021}. Following their work, in contrast to the slip models \citep{experimental_study_substrate_roughness_surfactant_effects_Landau_Levich_law_2005, liquid_film_entrainment_during_dip_coating_saturated_porous_substrate_2020}, the boundary condition we use considers both slip due to fluid patches at the equivalent surface (through $\bm{\mathcal{L}}$) and an excess tangential velocity driven by a pressure gradient over the rough layer (through $\bm{\mathcal{K}}^{itf}$). We employ a boundary condition of this form to model the thin film flow. Using the homogenization framework to compute $\bm{\mathcal{L}}$ and $\bm{\mathcal{K}}^{itf}$ for any surface with a periodic roughness pattern \citep{generalized_slip_condition_over_rough_surfaces_2019, interfacial_conditions_free_fluid_region_porous_medium_2021}, we can predict the coated film thickness without any fitting parameter.

We first present the solution for the coated film thickness using a homogenized boundary condition to model the rough surface (Section~\ref{sec:model}). Using the effective macroscopic properties of each surface derived from its microstructure, we solve the macroscopic lubrication equations and predict the coated film thickness. We present the experimental setup in Section~\ref{sec:experiments}, where we describe fabrication of surfaces having micropillars with varied geometric parameters and the interferometry technique used to measure $h_0$. The model is compared to experimental data (Section~\ref{sec:results}) and finally we discuss our results and their implications for further study of thin film flow over rough surfaces (Section~\ref{sec:discussionandconclusion}).

\section{Model of thin film flow over a porous bed}\label{sec:model}

The dip coating system is illustrated in figure~\ref{fig:dip_coating_system}, where a rough plate is pulled continuously from a liquid with density $\rho$, surface tension $\gamma$, and dynamic viscosity $\eta$. Three regions are observed: a flat film (Region I) is connected by a dynamic meniscus (Region II) to a static meniscus where the film meets the bath (Region III). We consider the film profile $h(z)$ indicated in figure~\ref{fig:dip_coating_system}(b) and, in particular, aim to find its limiting value $h \to h_0$ as $z \to \infty$.

We model the system by considering the flow of a thin liquid film over a rough or porous plate. The challenge in modeling such a system arises, first, because of the complex shape of the rough or porous surface, which is difficult to address by an analytical solution. The second challenge is the multiscale nature of the problem: the roughness amplitude $h_p$ may be much smaller than the film thickness $h_0$, which makes a direct numerical simulation computationally expensive. In addition, we desire to develop a general model for thin film flow over surfaces with arbitrary roughness topography, and we cannot arrive at such insight from seeking a separate numerical solution for each surface texture we wish to model. The homogenization method is an attractive choice for addressing these problems, because it can address arbitrarily complex periodic roughness patterns, yet requires only a full numerical solution within a periodic cell containing a single roughness feature to produce effective parameters for a simplified equivalent boundary condition \citep{permeability_rigid_porous_media_2010, generalized_slip_condition_over_rough_surfaces_2019, interfacial_conditions_free_fluid_region_porous_medium_2021}. These constant parameters are found by averaging the solution of associated Stokes problems in the microscopic domain \citep{generalized_slip_condition_over_rough_surfaces_2019, interfacial_conditions_free_fluid_region_porous_medium_2021}. The resulting boundary condition can be applied in an analytical approach to the full scale problem, which is treated in a similar manner as the classic analysis for a smooth plate  \citep{dragging_liquid_moving_plate_1942}. 

Before any simplification, our system consists of a Newtonian liquid flowing over a solid surface with arbitrary periodic roughness, which is governed by the Navier-Stokes equations \eqref{eq:full_Navier_Stokes} subject to the boundary condition that fluid cannot flow through the solid and that fluid velocity tangential to the solid is zero (no-slip). At the liquid-air interface, we apply the no-shear boundary condition and a capillary stress. The model proceeds in three steps: first, we present the governing equations in Section~\ref{subsec:governing_equations} and, after assuming a separation of scales between the texture's periodicity and the film thickness, in Section~\ref{subsec:boundary_conditions} we identify a homogenized boundary condition that represents the roughness of the plate yet can be applied at an equivalent flat surface $\mathbb{ES}$ (see figures~\ref{fig:boundary_condition_sketch}b, c) instead of the original complex roughness shape. Second, by assuming the film is thin, we perform a lubrication expansion to the governing equations and boundary conditions: the resulting simplified equations are presented in each respective section. Finally, in Section~\ref{subsec:solution} we solve the resulting equations for the film thickness $h_0$ coated on the plate as a function of velocity $v_0$ and the liquid properties ($\rho$, $\gamma$, $\eta$).

\subsection{Governing equations}
\label{subsec:governing_equations}
The rough surfaces are pulled from a Newtonian liquid of density $\rho$ and dynamic viscosity $\eta$, such that the flow satisfies the three-dimensional incompressible Navier-Stokes equations,
\begin{align}
    \frac{\partial \mathbf{u}}{\partial t} + (\mathbf{u} \cdot \nabla) \mathbf{u} &= \frac{\eta}{\rho} \nabla^2 \mathbf{u} - \frac{1}{\rho} \nabla p + \mathbf{g},
    \label{eq:full_Navier_Stokes} \\
    \nabla \cdot \mathbf{u} &= 0,
    \label{eq:mass_conservation}
\end{align}
where $\mathbf{u} = u_x \hat{\mathbf{x}} + u_y \hat{\mathbf{y}} + u_z \hat{\mathbf{z}}$ is the liquid velocity, $p$ is the pressure, and $\mathbf{g} = -g \hat{\mathbf{z}}$ ($g > 0$) is the gravitational acceleration. The operator $\nabla$ is defined as $\nabla = \hat{\mathbf{x}} \frac{\partial}{\partial x} + \hat{\mathbf{y}} \frac{\partial}{\partial y} + \hat{\mathbf{z}} \frac{\partial}{\partial z}$. In \eqref{eq:full_Navier_Stokes}, assuming that the system is in steady state, we drop the unsteady term, $\partial \mathbf{u}/\partial t = 0$. Assuming that velocities are zero in the $y$-direction, we simplify the problem to two dimensions $x$ and $z$ (see figure~\ref{fig:dip_coating_system}). We remark that while the assumption $u_y = 0$ is clearly sensible for a case of two-dimensional periodic structures that are invariant in the $y$-direction, such as ridges, is not obvious that velocities should be zero along $y$ for generic roughness patterns, such as a bed of pillars. Yet as long as the roughness pattern is periodic and isotropic along the $y$-axis, there is no reason for any symmetry breaking flow along $y$ and we may expect on average that that velocity is zero along $y$. Thus, for patterns that satisfy the condition of isotropy in the $y$-direction, this assumption is sensible.

Supposing that a characteristic film thickness scale $\ell'$ is much smaller than a characteristic scale $\ell''$ of the interface profile variation along the film surface, we perform a lubrication expansion in the small parameter $\delta = \ell'/\ell''$, arriving at the leading order equations \citep{reynolds1886, long_scale_evolution_thin_liquid_films_1997, pattern_formation_thin_liquid_films_2018}
\begin{align}
    \frac{\partial \tilde{p}}{\partial \tilde{x}} &= 0, 
    \label{eq:simplified_Navier_Stokes_x_nondim} \\
    \frac{\partial^2 \tilde{u}_z}{\partial \tilde{x}^2} &= \frac{\partial \tilde{p}}{\partial \tilde{z}} + \frac{\rho g \ell'^2}{\eta U_z},
    \label{eq:simplified_Navier_Stokes_z_nondim}
\end{align}
where $\tilde{\cdot}$ denotes dimensionless variables, and lengths and velocities have been normalized by characteristic scales
\begin{equation}
    x = \ell' \tilde{x}, \quad z = \ell'' \tilde{z}, \quad u_x = U_x \tilde{u}_x, \quad u_z = U_z \tilde{u}_z.
    \label{eq:characteristic_scales_lubrication}
\end{equation}
The pressure is normalized as
\begin{equation}
    p = \frac{\eta U_z}{\delta \ell'} \tilde{p} = \frac{\eta U_z}{\ell'^2/\ell''} \tilde{p}.
    \label{eq:pressure_lubrication}
\end{equation}
In addition, the liquid velocity normal to the plate is much smaller than its velocity parallel to the plate, so $U_x \sim \delta U_z$. This relation is required for mass conservation to hold at each order \citep{long_scale_evolution_thin_liquid_films_1997}. Later, we will require the dimensional form of \eqref{eq:simplified_Navier_Stokes_x_nondim}--\eqref{eq:simplified_Navier_Stokes_z_nondim}, which are given as
\begin{align}
    \frac{\partial p}{\partial x} &= 0,
    \label{eq:simplified_Navier_Stokes_x_dim} \\
    \frac{\eta}{\rho} \frac{\partial^2 u_z}{\partial x^2} &= \frac{1}{\rho} \frac{\partial p}{\partial z} + g.
    \label{eq:simplified_Navier_Stokes_z_dim}
\end{align}
Note that the characteristic scale $\ell''$ in our case is of the order of the capillary length $a$, which sets the meniscus scale in Region III (figure~\ref{fig:dip_coating_system}).

\subsection{Boundary conditions}
\label{subsec:boundary_conditions}

\subsubsection{Interface condition at the plate}
\label{subsubsec:interface_condition_at_plate}

\begin{figure}
    \centering
    \includegraphics[width=0.8\textwidth]{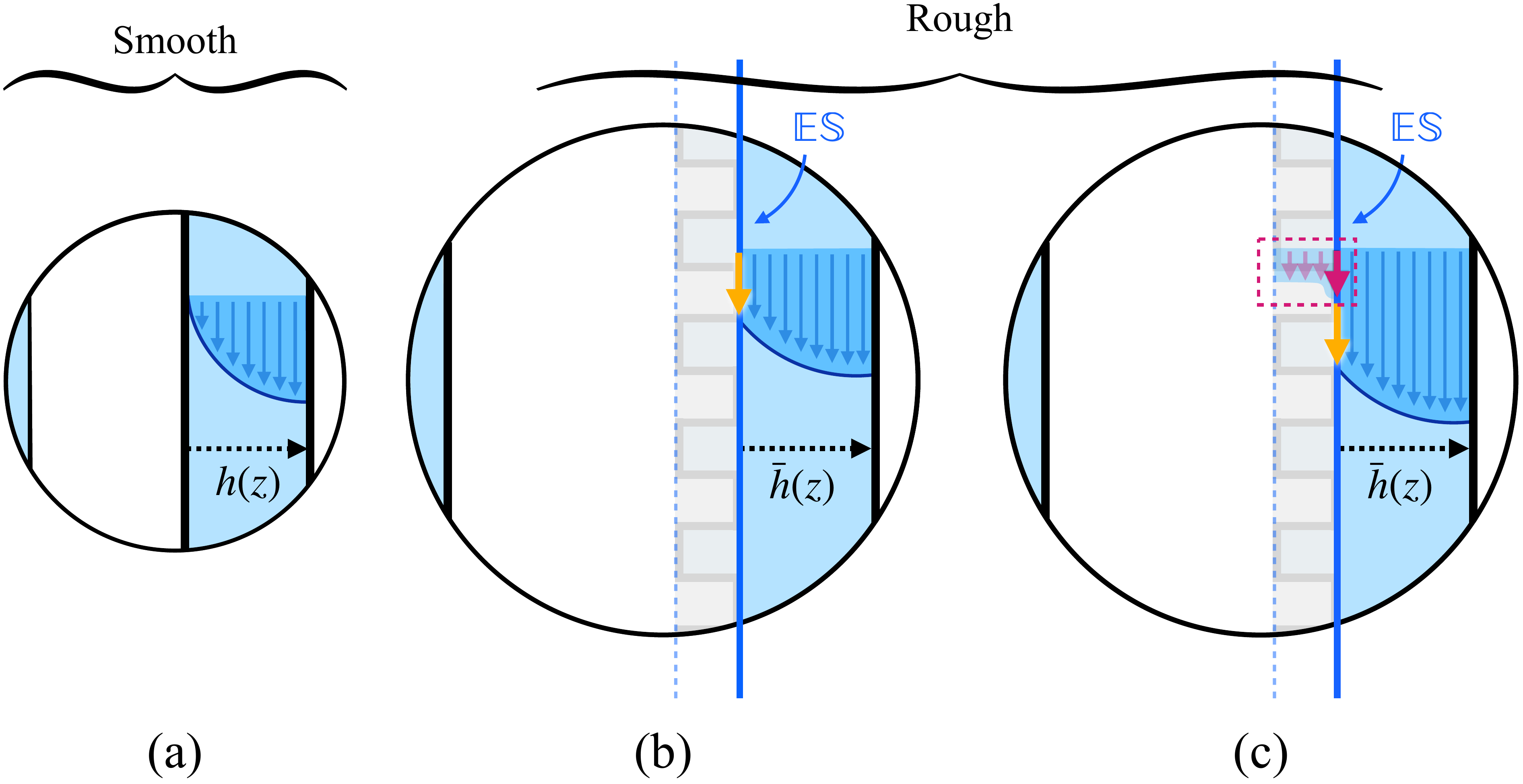}
    \caption{Boundary conditions for a smooth (a) and rough (b, c) plate. Velocity profiles are sketched in a reference frame moving with the plate. (a) No slip. (b) Slip. (c) Slip and porous flow. The yellow arrow represents the slip contribution and the pink arrow represents the contribution from porous flow, driven by a pressure gradient within the rough layer. For a rough surface, the slip contribution is $O(\epsilon^0)$ and the porous flow contribution is $O(\epsilon^1)$ \citep{interfacial_conditions_free_fluid_region_porous_medium_2021}.}
    \label{fig:boundary_condition_sketch}
\end{figure}

Supposing that the characteristic spacing $\ell$ between two periodic roughness elements is much smaller than the size of the liquid domain $\ell'$ ($\epsilon = \ell/\ell' \ll 1$), we use the homogenized boundary condition described in \citet{botNaq2020}, keeping terms up to $O(\epsilon^2)$. The parameter $\epsilon$ is a separation of scales parameter that is assumed to be small in the homogenization procedure. We first define a fictitious equivalent surface, denoted $\mathbb{ES}$, to be a flat plane located at a distance $d_{\mathbb{ES}}$ from the bottom of the roughness features (figure~\ref{fig:dip_coating_system}a). Similar to \citet{botNaq2020}, we select the location $d_{\mathbb{ES}} = h_p$ at the interface between the rough layer and the free fluid region (figures~\ref{fig:boundary_condition_sketch}b,c), based on the assumption that liquid is always trapped within the rough layer, as has been observed in prior experiments \citep{coating_textured_solid_2011}.\footnote{A detailed discussion regarding the position $d_{\mathbb{ES}}$ can be found in \citet{generalized_slip_condition_over_rough_surfaces_2019}.} In our case, $h_p$ represents a pillar height, but in general $h_p$ represents the characteristic peak-to-valley scale of the rough layer. Note that, due to the position of $\mathbb{ES}$, we will later decompose the total film profile $h(z)$ into two parts, $h(z) = h_p + \bar{h}(z)$, where $h_p$ is the known height of the trapped layer and $\bar{h}(z)$ is the unknown thickness of the free film above $\mathbb{ES}$: solving the macroscopic model will give the asymptotic \textit{free} film thickness $\bar{h}_0 \equiv \bar{h}(z \to\infty)$ measured with respect to $\mathbb{ES}$. The total film thickness will then be $h_0 = h_p + \bar{h}_0$. Having selected the equivalent surface's position, the boundary condition at $\mathbb{ES}$ is
\begin{align}
    u_x(x = 0, z) &= - \mathcal{K}^{itf} \frac{\partial}{\partial z} \left(\frac{\partial u_z}{\partial x} + \frac{\partial u_x}{\partial z}\right), 
    \label{eq:full_boundary_condition_ux_dim} \\
    u_z(x = 0, z) &= v_0 + \mathcal{L} \left(\frac{\partial u_z}{\partial x} + \frac{\partial u_x}{\partial z}\right) + \frac{\mathcal{K}^{itf}}{\eta} \frac{\partial}{\partial z} \left(-p + 2 \eta \frac{\partial u_x}{\partial x}\right),
    \label{eq:full_boundary_condition_uz_dim}
\end{align}
where $\mathcal{L}$ is the slip length and $\mathcal{K}^{itf}$ the interface permeability \citep{botNaq2020, interfacial_conditions_free_fluid_region_porous_medium_2021}. The conditions \eqref{eq:full_boundary_condition_ux_dim}--\eqref{eq:full_boundary_condition_uz_dim} on velocity $\mathbf{u}$ at the plate include the speed $v_0$ at which the plate is pulled in \eqref{eq:full_boundary_condition_uz_dim}, which by itself represents a typical no-slip condition, as well as the effect of the roughness pattern on the liquid velocity at the plate, represented by the $\mathcal{L}$ and $\mathcal{K}^{itf}$ terms. The form of the boundary conditions \eqref{eq:full_boundary_condition_ux_dim}--\eqref{eq:full_boundary_condition_uz_dim} indicates how roughness changes the boundary condition at $\mathbb{ES}$ as compared to the case of a smooth plate: the tangential velocity $u_z$ parallel to $\mathbb{ES}$ is augmented by a contribution of slip $\mathcal{L}$ and interface permeability $\mathcal{K}^{itf}$ and the velocity $u_x$ of liquid penetrating through $\mathbb{ES}$ can be nonzero, its value determined by $\mathcal{K}^{itf}$ \citep{botNaq2020, interfacial_conditions_free_fluid_region_porous_medium_2021}.

Applying the same scaling \eqref{eq:characteristic_scales_lubrication} as for the governing equations, we expand the boundary conditions \eqref{eq:full_boundary_condition_ux_dim}--\eqref{eq:full_boundary_condition_uz_dim} in $\delta$ and retrieve, at leading order in $\delta$,
\begin{align}
    u_x(x = 0, z) &= 0, 
    \label{eq:simplified_boundary_condition_ux_dim} \\
    u_z(x = 0, z) &= v_0 + \mathcal{L} \frac{\partial u_z}{\partial x} - \frac{\mathcal{K}^{itf}}{\eta} \frac{\partial p}{\partial z},
    \label{eq:simplified_boundary_condition_uz_dim}
\end{align}
The result of the lubrication expansion for $u_x$ in \eqref{eq:simplified_boundary_condition_ux_dim} indicates that at leading order in $\delta$, velocities perpendicular to the equivalent surface are zero. Thus, in the model, no fluid transfers from the free-fluid region to the rough layer. This result is sensible in light of findings that the wall-normal velocity $u_x$ at $\mathbb{ES}$ is typically one to two orders of magnitude smaller than the tangential slip velocity \citep{higher_order_homogenized_boundary_conditions_flows_rough_porous_surfaces_2021}.

The boundary condition \eqref{eq:simplified_boundary_condition_uz_dim} for flow tangential to $\mathbb{ES}$ is illustrated in figure~\ref{fig:boundary_condition_sketch}(c), where it is compared to the typical no-slip condition (figure~\ref{fig:boundary_condition_sketch}a) and Navier slip condition (figure~\ref{fig:boundary_condition_sketch}b). Velocity profiles have been drawn in the reference frame of the moving plate. As illustrated in figure~\ref{fig:boundary_condition_sketch}(b), the slip condition modifies the velocity profile by augmenting the velocity at $\mathbb{ES}$ (the $\mathcal{L} \frac{\partial u_z}{\partial x}$ term). When we do not neglect the flow through the rough layer, the slip contribution is augmented further by a flow along the interface driven by the pressure gradient within and above the rough layer (the $\frac{\mathcal{K}^{itf}}{\eta} \frac{\partial p}{\partial z}$ term), illustrated as the pink arrow in figure~\ref{fig:boundary_condition_sketch}(c). Even if the slip and permeability contribution in \eqref{eq:simplified_boundary_condition_uz_dim} are at the same order in terms of the lubrication parameter $\delta$, one should note that these contributions are not equally important in terms of the homogenization parameter $\epsilon$ \citep{interfacial_conditions_free_fluid_region_porous_medium_2021}. The interface permeability term is usually found to be negligible ($O(\epsilon^1)$) compared to the slip term ($O(\epsilon^0)$) in the case of laminar flow over rough surfaces in an unbounded fluid domain \citep{generalized_slip_condition_over_rough_surfaces_2019}, but, as we will see, this term cannot be neglected in the case of a thin film flow. Note that the interface permeability term is absent in previously proposed slip models \citep{experimental_study_substrate_roughness_surfactant_effects_Landau_Levich_law_2005, coating_textured_solid_2011}.

\subsubsection{Interface condition at the liquid-air interface}
\label{subsubsec:interface_condition_at_free_surface}

At leading order in $\delta$, we impose at the free surface a zero shear stress condition,
\begin{equation}
    \eta \frac{\partial u_z}{\partial x}(x = \bar{h}, z) = 0,
    \label{eq:no_tangential_stress_boundary_condition}
\end{equation}
as well as capillary pressure at the free interface, expressed as
\begin{equation}
    p(x = \bar{h}, z) = -\gamma \frac{\frac{\partial^2 \bar{h}}{\partial z^2}}{\left[1 + \left(\frac{\partial \bar{h}}{\partial z}\right)^2\right]^{3/2}} \approx -\gamma \frac{\partial^2 \bar{h}}{\partial z^2}.
    \label{eq:pressure_free_surface}
\end{equation}
where the final expression for interface curvature assumes small slopes \citep{dragging_liquid_moving_plate_1942, pattern_formation_thin_liquid_films_2018}.

\subsection{Solution}
\label{subsec:solution}

To find $h_0$ for a given surface, we solve the system of governing equations \eqref{eq:simplified_Navier_Stokes_x_dim}--\eqref{eq:simplified_Navier_Stokes_z_dim} combined with the boundary conditions \eqref{eq:simplified_boundary_condition_ux_dim}--\eqref{eq:simplified_boundary_condition_uz_dim}. In the next section, we will compare the results from considering two cases for the boundary condition \eqref{eq:simplified_boundary_condition_uz_dim}: a case where only the slip term is present and a case where flow through the rough layer also contributes.

We begin by finding the velocity profile in the thin film. After integrating \eqref{eq:simplified_Navier_Stokes_z_dim} once, we arrive at
\begin{equation}
    \frac{\partial u_z}{\partial x} = \left(\frac{1}{\eta} \frac{\partial p}{\partial z} + \frac{\rho g}{\eta}\right) x + F(z),
    \label{eq:first_indefinite_integral}
\end{equation}
and integrate a second time to arrive at the velocity profile
\begin{equation}
    u_z = \left(\frac{1}{\eta} \frac{\partial p}{\partial z} + \frac{\rho g}{\eta}\right) \frac{x^2}{2} + F(z) x + G(z),
    \label{eq:second_indefinite_integral}
\end{equation}
where $F$ and $G$ are unknown functions of $z$. Applying the no-stress condition \eqref{eq:no_tangential_stress_boundary_condition} at the liquid-air interface $x = \bar{h}$, we solve for $F(z)$ using \eqref{eq:first_indefinite_integral},
\begin{equation}
    F(z) = -\left(\frac{1}{\eta} \frac{\partial p}{\partial z} + \frac{\rho g}{\eta}\right) \bar{h}.
\end{equation}
At the solid-liquid interface $x = 0$, substitute the boundary condition \eqref{eq:simplified_boundary_condition_uz_dim} into \eqref{eq:second_indefinite_integral} to solve for $G(z)$,
\begin{align}
    u_z(x = 0, z) = G(z) &= v_0 + \mathcal{L} \frac{\partial u_z}{\partial x} - \frac{\mathcal{K}^{itf}}{\eta} \frac{\partial p}{\partial z}
    \nonumber \\
    &= v_0 - \mathcal{L} \left(\frac{1}{\eta} \frac{\partial p}{\partial z} + \frac{\rho g}{\eta}\right) \bar{h} - \frac{\mathcal{K}^{itf}}{\eta} \frac{\partial p}{\partial z},
    \label{eq:velocity_profile}
\end{align}
where we know that $\frac{\partial p}{\partial z}(x = 0, z) = \frac{\partial p}{\partial z}(z)$ for all $x$, because $p$ does not depend on $x$ by \eqref{eq:simplified_Navier_Stokes_x_dim}. From the form of $G(z)$ in \eqref{eq:velocity_profile}, which gives the velocity profile \eqref{eq:second_indefinite_integral} at $x = 0$, it is apparent that the boundary condition at the rough interface acts as a correction to the no-slip boundary condition used in the dip coating problem for a smooth surface, in which $G(z) = v_0$ \citep{dragging_liquid_moving_plate_1942}. The velocity profile in the thin film is thus
\begin{equation}
    u_z(x, z) = \left(\frac{1}{\eta} \frac{\partial p}{\partial z} + \frac{\rho g}{\eta}\right) \left(\frac{x^2}{2} - \bar{h} x - \bar{h}\mathcal{L}\right) - \frac{\mathcal{K}^{itf}}{\eta} \frac{\partial p}{\partial z} + v_0, 
\end{equation}
and, substituting the pressure from \eqref{eq:pressure_free_surface}, we have
\begin{equation}
    u_z(x, z) = \left(\frac{\rho g}{\eta} - \frac{\gamma}{\eta} \frac{d^3\bar{h}}{dz^3}\right) \left(\frac{x^2}{2} - \bar{h} x - \bar{h}\mathcal{L}\right) + \frac{\gamma}{\eta} \frac{d^3\bar{h}}{dz^3} \mathcal{K}^{itf} + v_0.
    \label{eq:velocity_profile_substituted_pressure}
\end{equation}

For an incompressible fluid, the flux $j$ per unit plate width can be written
\begin{equation}
    j = \int_0^{\bar{h}} u_z dx = \text{constant.}
    \label{eq:flux_definition}
\end{equation}
Integrating $u_z$ \eqref{eq:velocity_profile_substituted_pressure} to compute $j$ \eqref{eq:flux_definition} produces
\begin{equation}
    j = v_0 \bar{h} + \left(\frac{\rho g}{\eta} - \frac{\gamma}{\eta} \frac{d^3 \bar{h}}{dz^3} \right) \left(- \frac{\bar{h}^3}{3} - \mathcal{L} \bar{h}^2 \right) + \frac{\gamma}{\eta} \frac{d^3 \bar{h}}{dz^3} \mathcal{K}^{itf} \bar{h}.
    \label{eq:flux_j_porous_case}
\end{equation}
Keeping in mind that $j$ is a constant, \eqref{eq:flux_j_porous_case} defines the liquid layer thickness $\bar{h} = \bar{h}(z)$. We rewrite the ordinary differential equation \eqref{eq:flux_j_porous_case} for $\bar{h}$ as
\begin{equation}
    \frac{d^3 \bar{h}}{dz^3} = \frac{3\eta}{\gamma} \frac{(j - v_0 \bar{h})}{{\bar{h}}^3 + 3 \mathcal{L} {\bar{h}}^2 + 3 \mathcal{K}^{itf} \bar{h}} + \frac{\rho g}{\gamma} \frac{{\bar{h}}^3 + 3 \mathcal{L} {\bar{h}}^2}{{\bar{h}}^3 + 3 \mathcal{L} {\bar{h}}^2 + 3 \mathcal{K}^{itf} \bar{h}}.
    \label{eq:ordinary_differential_equation_for_film_thickness_h}
\end{equation}
To solve the third-order differential equation for $\bar{h}$, we require three boundary conditions, which will come from matching the dynamic meniscus profile $\bar{h}(z)$ to the flat film (figure~\ref{fig:dip_coating_system}, Region I). Prior to that, we nondimensionalize \eqref{eq:ordinary_differential_equation_for_film_thickness_h} \citep{dragging_liquid_moving_plate_1942}. Introducing a nondimensional film thickness $\mu$, defined as
\begin{equation}
    \mu = \frac{v_0 \bar{h}}{j},
    \label{eq:nondimensional_film_height}
\end{equation}
we can rearrange \eqref{eq:ordinary_differential_equation_for_film_thickness_h} to arrive at
\begin{equation}
    \frac{d^3 \mu}{dz^3} = \frac{3\eta}{\gamma} \frac{v_0^4}{j^3} \frac{(1 - \mu)}{\mu \left(\mu^2 + \frac{3v_0}{j} \mathcal{L} \mu + \frac{3 v_0^2}{j^2} \mathcal{K}^{itf}\right)} + \frac{v_0}{j} \frac{\rho g}{\gamma} \frac{\mu \left( \mu^2 + \frac{3v_0}{j} \mathcal{L} \mu \right)}{\mu \left(\mu^2 + \frac{3v_0}{j} \mathcal{L} \mu + \frac{3v_0^2}{j^2} \mathcal{K}^{itf} \right)}.
\end{equation}
The scale $3\eta v_0^4 / (\gamma j^3)$ naturally provides a change of variables for the spatial coordinate $z$, so along the plate we introduce the nondimensional spatial coordinate $\lambda$, defined as
\begin{equation}
    \lambda = \left(\frac{3\eta}{\gamma}\right)^{1/3} \frac{v_0^{4/3}}{j} z,
\end{equation}
and arrive at a nondimensional version of equation \eqref{eq:ordinary_differential_equation_for_film_thickness_h},
\begin{equation}
    \frac{d^3 \mu}{d\lambda^3} = \frac{(1 - \mu)}{\mu \left(\mu^2 + \mathcal{L}^* \mu + K^*\right)} + \frac{\rho g j^2}{3\eta v_0^3} \frac{\mu \left( \mu^2 + \mathcal{L}^* \mu \right)}{\mu \left(\mu^2 + \mathcal{L}^* \mu + K^* \right)},
\end{equation}
where we have defined a nondimensional slip $\mathcal{L}^*$ and interface permeability $K^*$,
\begin{equation}
    \mathcal{L}^* = \frac{3 v_0}{j} \mathcal{L}, \quad K^* = \frac{3 v_0^2}{j^2} \mathcal{K}^{itf}.
    \label{eq:nondimensional_slip_and_permeability}
\end{equation}

Assuming that $\rho g j^2 / (3 \eta v_0^3) \ll 1$, which is valid for low pulling velocities $Ca \to 0$ \citep{dragging_liquid_moving_plate_1942, drag_out_problem_film_coating_theory_1982}, we neglect the final term, giving the simplified equation
\begin{equation}
    \frac{d^3 \mu}{d\lambda^3} = \frac{(1 - \mu)}{\mu \left(\mu^2 + \mathcal{L}^* \mu + K^* \right)},
    \label{eq:differential_equation_film_thickness_mu_simplified}
\end{equation}
relating film thickness $\mu$ to position $\lambda$ along the plate. Following \cite{dragging_liquid_moving_plate_1942}, we conclude that $\mu \to 1$ in the flat film region, 
\begin{equation}
    \lim_{\lambda\to\infty} \mu = 1,
    \label{eq:flat_film_limit}
\end{equation}
since higher derivatives of $\mu$ must tend to zero if the film is flat. In other words, we have $\frac{d^3 \mu}{d\lambda^3} \to 0$ in \eqref{eq:differential_equation_film_thickness_mu_simplified}. Applying a small perturbation $\mu = 1 + \mu_1$, and solving the resulting equations at the leading order produces the matching conditions in the dynamic meniscus region as $\lambda \to \infty$,
\begin{align}
    \mu(\lambda \to \infty) &= 1 + A e^{-\frac{\lambda}{\sqrt[3]{1 + \mathcal{L}^* + K^*}}},
    \label{eq:boundary_condition_mu} \\
    \frac{d\mu}{d\lambda}(\lambda \to \infty) &= (1 - \mu) \left[1 + \mathcal{L}^* + K^*\right]^{-\frac{1}{3}}, 
    \label{eq:boundary_condition_dmu_dlambda} \\
    \frac{d^2\mu}{d\lambda^2}(\lambda \to \infty) &= (\mu - 1) \left[1 + \mathcal{L}^* + K^*\right]^{-\frac{2}{3}}.
    \label{eq:boundary_condition_d2mu_dlambda2}
\end{align}
where $A$ is an unknown constant of integration. The constant $A$ poses a problem, yet one that can be overcome by the variable transformation used in \cite{dragging_liquid_moving_plate_1942}, one which is also favorable for numerical integration, because it transforms the boundary location to a finite value rather than $\lambda \to \infty$. To numerically solve the film thickness equation \eqref{eq:differential_equation_film_thickness_mu_simplified} together with boundary conditions \eqref{eq:boundary_condition_mu}--\eqref{eq:boundary_condition_d2mu_dlambda2}, we define the transformation (see Appendix~\ref{appendix:transformation})
\begin{equation}
    \xi = \left(\frac{d\mu}{d\lambda}\right)^2,
    \label{eq:lambda_to_xi_transformation}
\end{equation}
and arrive at the differential equation
\begin{equation}
    \frac{d^2\xi}{d\mu^2} = \frac{2 (\mu - 1)}{\mu \left(\mu^2 + \mathcal{L}^* \mu + K^* \right) \sqrt{\xi}}
    \label{eq:differential_equation_xi_mu}
\end{equation}
with boundary conditions
\begin{align}
    \xi(\mu \to 1) &= (1 - \mu)^2 \left[1 + \mathcal{L}^* + K^*\right]^{-\frac{2}{3}},
    \label{eq:boundary_condition_xi} \\
    \frac{d\xi}{d\mu}(\mu \to 1) &= 2(\mu - 1) \left[1 + \mathcal{L}^* + K^*\right]^{-\frac{2}{3}},
    \label{eq:boundary_condition_dxi_dmu}
\end{align}
where we now apply the boundary condition at a finite value $\mu \to 1$, since we know $\mu \to 1$ as $\lambda \to \infty$. Note that \eqref{eq:lambda_to_xi_transformation} implies a transformation of the film curvature as
\begin{align}
    \frac{d^2\mu}{d\lambda^2} &= \frac{1}{2} \frac{d\xi}{d\mu},
    \label{eq:d2mu_dlambda2_transformed}
\end{align}
which will be important later. Reducing the order of the equations from third- to second-order has the consequence that we will not be able to solve exactly for the full film thickness profile $\mu(\lambda)$, since the unknown constant $A$ remains. However, as we will see, knowing that the film curvature $\frac{d^2\mu}{d\lambda^2}$ is given by \eqref{eq:d2mu_dlambda2_transformed} will be sufficient to find the film thickness in the flat film region where $\mu(\lambda \to \infty) = 1$ or, equivalently, where $\bar{h}_0 \equiv \bar{h}(z \to \infty) = j / v_0$ (see \eqref{eq:nondimensional_film_height}).

The system of equations \eqref{eq:differential_equation_xi_mu} with \eqref{eq:boundary_condition_xi}--\eqref{eq:boundary_condition_dxi_dmu} is not closed, because the unknown flux $j$ remains in the terms $\mathcal{L}^* = \mathcal{L}^*(j)$ and $\mathcal{K}^* = K^*(j)$ (see \eqref{eq:nondimensional_slip_and_permeability}). We know that the flux $j$ is governed by the flow of liquid through the meniscus region where the flat film meets the bath, because there the capillary suction competes with viscous stress to determine $h_0$. Thus, we can find $j$ by considering the form of the static meniscus in Region III (figure~\ref{fig:dip_coating_system}). Assuming that the meniscus is quasi-static, we solve equations balancing capillary and hydrostatic pressure to find the static meniscus shape and arrive at a matching condition for the meniscus curvature at the bath \citep{dragging_liquid_moving_plate_1942}
\begin{equation}
    \frac{d^2\bar{h}}{dz^2}(z \to 0) = \frac{\sqrt{2}}{a}, 
    \label{eq:meniscus_curvature_dimensional}
\end{equation}
where we recall that $a = \sqrt{\gamma/(\rho g)}$ is the capillary length. Nondimensionalizing from $(z, \bar{h})$ to $(\lambda, \mu)$ using \eqref{eq:nondimensional_film_height}, we have the curvature condition for the bottom of the dynamic meniscus,
\begin{equation}
    \frac{d^2\mu}{d\lambda^2}(\lambda \to 0) = \frac{\sqrt{2} \gamma^{2/3} j}{a v_0^{5/3} (3\eta)^{2/3}}, 
    \label{eq:meniscus_curvature_nondimensional}
\end{equation}
now in terms of the unknown flux $j$ we seek. Defining the meniscus curvature
\begin{equation}
    \kappa \equiv \frac{d^2\mu}{d\lambda^2}(\lambda \to 0),
    \label{eq:kappa_definition}
\end{equation}
we see that the dynamic meniscus governed by \eqref{eq:differential_equation_film_thickness_mu_simplified} or, equivalently, \eqref{eq:differential_equation_xi_mu}, must exhibit a profile $\mu(\lambda)$ whose curvature $\kappa = \frac{d^2\mu}{d\lambda^2}(\lambda \to 0)$ matches the value on the right-hand side of \eqref{eq:meniscus_curvature_nondimensional}. The left-hand side of \eqref{eq:meniscus_curvature_nondimensional} can be solved for using \eqref{eq:differential_equation_xi_mu} and the transformation \eqref{eq:d2mu_dlambda2_transformed}, yet its solution still depends on the value of the unknown $j$ because $\mathcal{L}^* = \mathcal{L}^*(j)$ and $K^* = K^*(j)$. The right-hand side is also determined except for the unknown $j$. Putting aside for now that $j$ is unknown, knowing that $\mu \to 1$ in the flat film region as $\lambda \to \infty$ (see \eqref{eq:flat_film_limit}), from \eqref{eq:meniscus_curvature_nondimensional} we have an expression for the free film thickness $\bar{h} \to \bar{h}_0$ where
\begin{equation}
    \bar{h}_0 = \frac{j}{v_0} = \frac{\kappa}{\sqrt{2}} \frac{v_0^{2/3} (3\eta)^{2/3}}{\gamma^{1/6} (\rho g)^{1/2}}. 
    \label{eq:free_film_thickness_solution}
\end{equation}
The last equality comes from using the expression for $j$ derived in \eqref{eq:meniscus_curvature_nondimensional}, where it becomes clear how matching to the static meniscus curvature provides the information needed to determine $\bar{h}_0$. In terms of capillary length $a$ and capillary number $Ca$, \eqref{eq:free_film_thickness_solution} becomes
\begin{equation}
    \bar{h}_0 = \frac{3^{2/3} \kappa}{\sqrt{2}} a Ca^{2/3},
    \label{eq:free_film_thickness_solution_Ca_and_a}
\end{equation}
which has the form of \eqref{eq:smooth_surface_film_thickness}, except that the curvature $\kappa = \kappa(\mathcal{L}, \mathcal{K}^{itf})$ is no longer constant but a function of the slip $\mathcal{L}$ and interface permeability $\mathcal{K}^{itf}$.

To determine the value of the curvature and close the solution \eqref{eq:free_film_thickness_solution_Ca_and_a}, we examine the equation for $\kappa$ given in \eqref{eq:meniscus_curvature_nondimensional}: after rearranging the right-hand side and defining the nondimensional parameter $H = \bar{h}_0 / a =  j/(v_0 a)$, we notice that \eqref{eq:meniscus_curvature_nondimensional} has the form
\begin{equation}
    \kappa(H) = \frac{\sqrt{2}}{3^{2/3}} Ca^{-2/3} H.
    \label{eq:fixed_point_kappa}
\end{equation}
For completeness, we provide the definition of $\kappa(H)$ \eqref{eq:kappa_definition} in terms of the transformation \eqref{eq:d2mu_dlambda2_transformed} as
\begin{equation}
    \kappa(H) \equiv \frac{d^2\mu}{d\lambda^2}(\lambda \to 0) = \frac{1}{2} \frac{d\xi}{d\mu}(\mu \to \infty) = \lim_{\mu \to \infty} \int \frac{\mu - 1}{\mu (\mu^2 + \mathcal{L}^*(H) \mu + K^*(H)) \sqrt{\xi}} d\mu,
\end{equation}
which makes clear the dependence of $\kappa$ on the unknown flux $j$, or, equivalently, the unknown parameter $H$. Note that $\mathcal{L}^*(H) = 3\mathcal{L}/(Ha)$, and similarly $K^*(H) = 3\mathcal{K}^{itf}/(Ha)^2$.

A nonlinear equation of the form \eqref{eq:fixed_point_kappa} can be solved numerically by fixed-point iteration \citep{numerical_analysis_2011}: 

(i) We first provide an initial guess $H_0$ to solve for $\kappa_0 = \kappa(H_0)$. The solution for $\kappa$ is retrieved by solving \eqref{eq:differential_equation_xi_mu}--\eqref{eq:boundary_condition_dxi_dmu} and using the transformation \eqref{eq:d2mu_dlambda2_transformed} to compute $\kappa = \frac{d^2\mu}{d\lambda^2}(\lambda \to 0)$.

(ii) For $i > 0$, compute $H_i = \frac{3^{2/3}}{\sqrt{2}} Ca^{2/3} \kappa_{i-1}$ and solve for $\kappa_i = \kappa(H_i)$.

(iii) If the difference $\Delta = (H_i - H_{i-1})/H_{i-1}$ is acceptably small, stop the iteration. Otherwise, continue the iteration. Here we stop the iteration using a relative tolerance value $\Delta \leq 10^{-6}$.

\noindent The iteration is implemented in Python, where we solve the equations \eqref{eq:differential_equation_xi_mu}--\eqref{eq:boundary_condition_dxi_dmu} for $\kappa$ using the SciPy package \citep{scipy_fundamental_algorithms_scientific_computing_python_2020}. Code is provided in the Supplementary Material. Knowing the nondimensional free film thickness $H$ from the fixed-point iteration, we immediately arrive at the dimensional free film thickness $\bar{h}_0 = Ha$, or, equivalently, the fixed-point iteration provides the solution for $\kappa$, meaning that the expressions \eqref{eq:free_film_thickness_solution} and \eqref{eq:free_film_thickness_solution_Ca_and_a} are fully determined.

\section{Dip coating experiments}\label{sec:experiments}

\begin{figure}
    \centering
    \includegraphics[width=\textwidth]{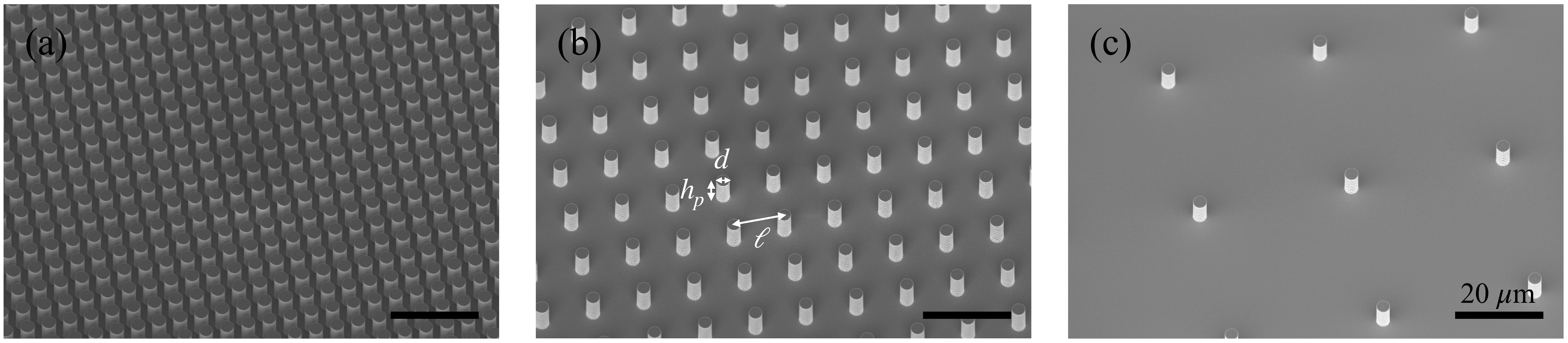}
    \caption{Rough surfaces etched into silicon wafers. In these three examples, pillars have height $h_p = 7.2 \pm 0.2$~\textmu m, diameter $d = 3.3 \pm 0.1$ \textmu m, and spacings (a) $\ell = 5$~\textmu m, (b) $\ell = 12$~\textmu m, and (c) $\ell = 36$~\textmu m. Scale bars are 20~\textmu m.}
    \label{fig:surfaces}
\end{figure}

\begin{table}
    \centering
    \def~{\hphantom{0}}
    \begin{tabular}{cccccc}
        $\phi$ (\%) & $\ell$ (\textmu m) & $\mathcal{L}$ (\textmu m) & $\mathcal{L}/\ell$ & $\mathcal{K}^{itf}$ (\textmu m$^2$) & $\mathcal{K}^{itf}/\ell^2$ \\ \hline
        0.2 & 72 & 6.84 & 0.095 & 26.96 & 0.0052 \\
        0.7 & 36 & 5.94 & 0.165 & 20.61 & 0.0159 \\
        2.6 & 18 & 3.55 & 0.197 & 10.82 & 0.0334 \\
        6 & 12 & 1.91 & 0.159 & 4.54 & 0.0315 \\
        24 & 6 & 0.33 & 0.055 & 0.25 & 0.0069 \\
        34 & 5 & 0.16 & 0.031 & 0.07 & 0.0026 \\
    \end{tabular}
    \caption{Six rough surfaces are used for the experiments, having pillars with constant diameter $d = 3.3 \pm 0.1$ \textmu m, constant height $h_p = 7.2 \pm 0.2$ \textmu m, and varied spacing $s$. The surface fraction is $\phi = \pi d^2/(4\ell^2)$. Computed slip and interface permeability values used in the model (see Section~\ref{sec:model} and Appendix~\ref{appendix:microscopic_problem}) are listed in dimensional ($\mathcal{L}$,~$\mathcal{K}^{itf}$) and nondimensional ($\mathcal{L}/\ell$,~$\mathcal{K}^{itf}/\ell^2$) forms.}
    \label{tab:surfaces}
\end{table}

To probe how roughness modifies the coated film thickness for varied roughness parameters, we perform experiments in which a viscous silicone oil coats rough silicon wafers. Rough surfaces are etched from silicon using photolithography and dry etching to produce a square grid of micropillars (figure~\ref{fig:surfaces}). Pillar height is $h_p = 7.2 \pm 0.2$ \textmu m and pillar diameter is $d = 3.3 \pm 0.1$ \textmu m. Spacing between the pillars varies between $\ell = 5-72$ \textmu m, so that the solid area fraction varies between $\phi = 0.3-50\%$. Parameters for the surfaces tested are listed in table \ref{tab:surfaces}.

\begin{figure}
    \centering
    \includegraphics[width=0.95\textwidth]{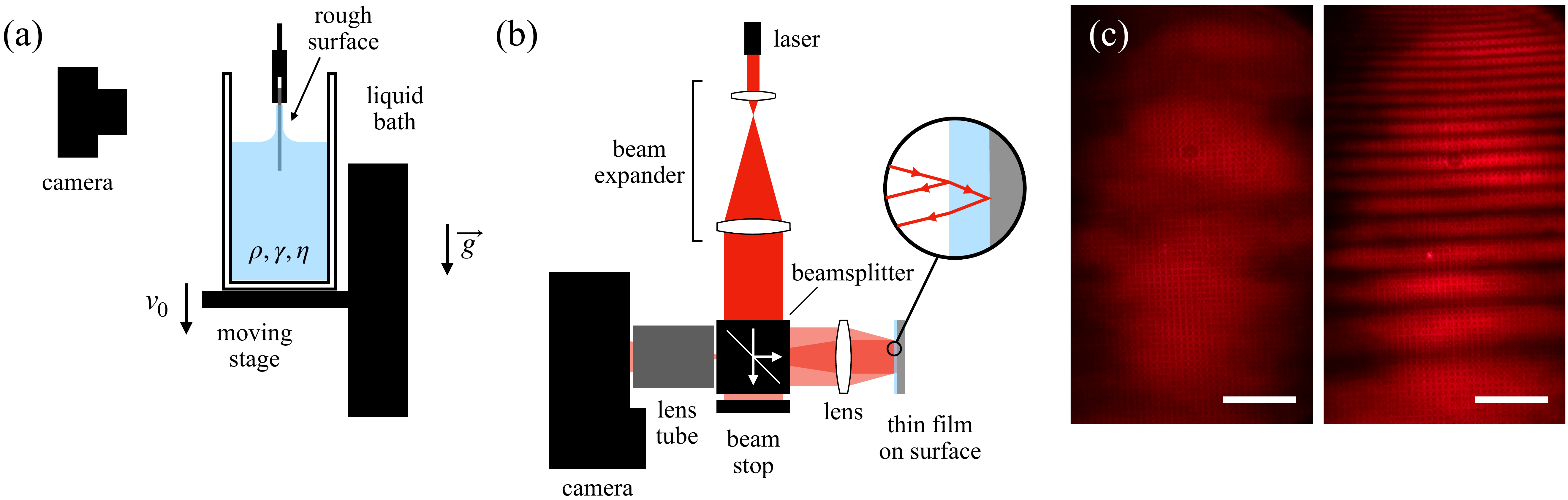}
    \caption{Dip coating experimental apparatus. (a) Side-view. A rough surface is held in place while a liquid bath of density $\rho$, surface tension $\gamma$, and dynamic viscosity $\eta$ moves downward at speed $v_0$. A camera records an interferometric image of the experiment. (b) Top-view. A laser passes through a beam expander and into a beamsplitter which directs it toward the thin film of liquid. The light interferes in the thin film and an image of the interference pattern is recorded by the camera. (c) A typical interfererometric image of a thin film of silicone oil on a rough silicon wafer \textit{(left)} during the steady regime and \textit{(right)} during the drainage regime after the bath has stopped moving. Scale bars are 0.5 mm.}
    \label{fig:experimental_apparatus}
\end{figure}

The dip coating experimental apparatus is illustrated in figure~\ref{fig:experimental_apparatus}. A bath of dimensions $9.0$~cm $\times$ $9.0$~cm $\times$ $19.0$~cm (length $\times$ width $\times$ depth) is constructed from acrylic plates and filled with silicone oil (Silitech AG) of density $\rho = 941 \pm 2$~kg/m$^3$, surface tension $\gamma = 21.2 \pm 0.2$~mN/m, and viscosity $\eta = 20.175 \pm 0.005$~mPa~s, where the measured values are reported with 95\% confidence intervals. The bath's length/width ($9.0$~cm $\approx$ $60a$) and depth ($19.0$~cm $\approx$ $125a$) are many times larger than the capillary length $a \approx 1.5$ mm, such that the edges of the bath do not affect the flow \citep{landau_levich_flow_visualization_2012, confinement_effects_dip_coating_2017}. The solid plates are approximately 5~cm $\times$ 2~cm (length $\times$ width) and 0.05~cm thick. The surfaces are wide enough ($2$~cm) that flow near the edges of the plate does not modify film thickness near the center where the measurement is taken \citep{self_similar_draining_near_vertical_edge_2020}, which is verified visually and by confirming that measured film thicknesses match predicted values in the case of a smooth plate. The bath is placed on a stage that can be moved at speeds between $v_0 = 1$ \textmu m/s -- $5$ mm/s by a linear motor (CONEX-LTA-HS, Newport).

A rough surface is held stationary in the liquid and the bath is moved downward at a constant velocity $v_0$, while a camera (Nikon D850) records an interferometric image of the film (figure~\ref{fig:experimental_apparatus}a). The optical path is illustrated in figure~\ref{fig:experimental_apparatus}b. A red laser (Arima ADL-63054TL) with wavelength $\lambda_0 = 635$ nm is directed toward the thin liquid film and reflected back towards a camera (Nikon D850), where the measured intensity depends on the film thickness due to thin film interference \citep{practical_realisation_length_by_interferometry_2018}. When the experiment is running, the bath has constant velocity $v_0$ relative to the plate, the film thickness is roughly constant, and thus the intensity is fairly constant across the image (figure~\ref{fig:experimental_apparatus}c, left panel). After the bath stops moving downward, we reach the drainage regime \citep{draining_vertical_plate_1930, entrainments_visqueux_2010} and the film begins to thin, producing a stripe pattern (figure~\ref{fig:experimental_apparatus}c, right panel). 

To obtain a reference height, and thus translate the number of fringes to an absolute film thickness \citep{practical_realisation_length_by_interferometry_2018}, we record the intensity during the film drainage period and finally move the field of view until we see the fringe where the free film meets the trapped film; this reference fringe thus provides a reference for $\bar{h}(z) = 0$. Material characterization and representative experimental videos may be found in the Supplementary Material.

\section{Results}\label{sec:results}

Having solved for the free film thickness $\bar{h}_0$ \eqref{eq:free_film_thickness_solution_Ca_and_a} as a function of meniscus curvature $\kappa$, capillary length $a$, and capillary number $Ca$, we are able to compute $h_0$ for varied surface roughness (given by particular $\mathcal{L}$ and $\mathcal{K}^{itf}$ values), liquids (encoded within $a$ and $Ca$), and dip coating velocities (given by $Ca$). In Sections~\ref{subsec:slip_and_interface_perm_relation}~and~\ref{subsec:model_predictions}, we solve the microscopic problem to compute effective parameters $\mathcal{L}$ and $\mathcal{K}^{itf}$, and demonstrate how these parameters determine macroscopic model predictions. In Section~\ref{subsec:comparison_with_experiments}, we compare our model to experimental data.

\subsection{Microtexture slip $\mathcal{L}$ and interface permeability $\mathcal{K}^{itf}$}\label{subsec:slip_and_interface_perm_relation}

\begin{figure}
    \centering
    \includegraphics[scale=0.65]{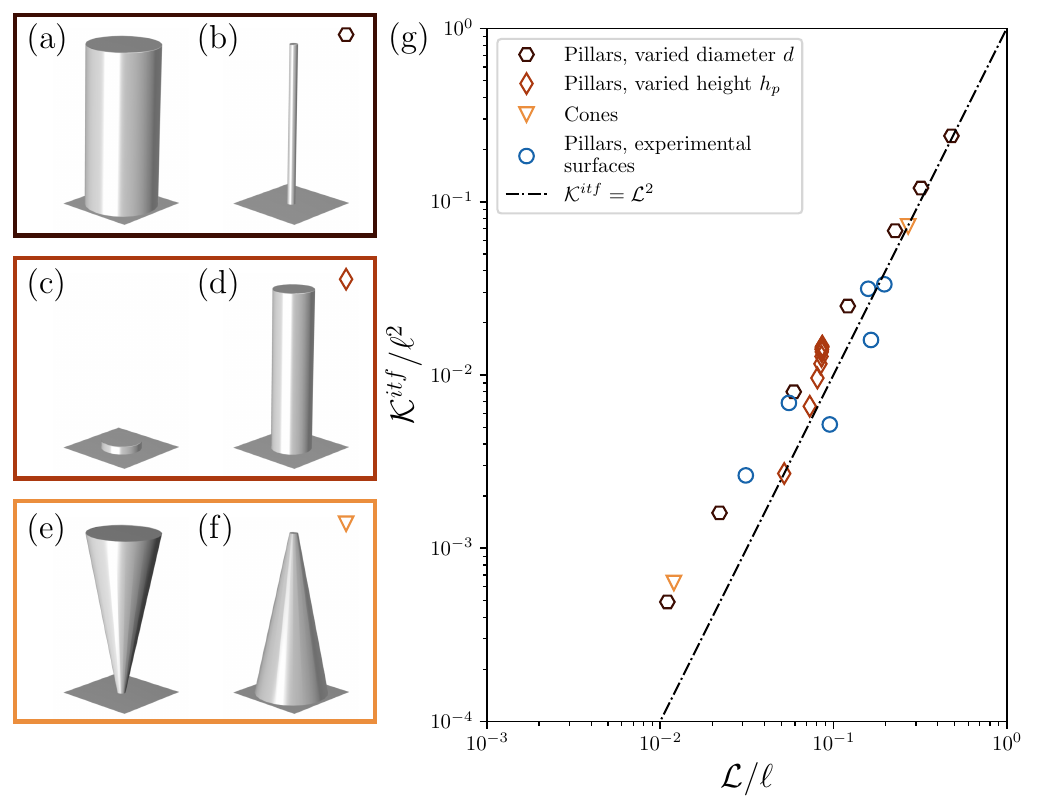}
    \caption{Nondimensional slip $\mathcal{L}/\ell$ and interface permeability $\mathcal{K}^{itf}/\ell^2$ for different surface designs (g), together with the relation $\mathcal{K}^{itf} = \mathcal{L}^2$ as a dash-dotted line \citep{boundary_conditions_naturally_permeable_wall_1967}. The surfaces were periodic with the following unit structures: pillars with diameter $d$ varying from large (a) to small (b) with fixed height $h_p$ and spacing $\ell$, pillars with $h_p$ varying from (c) small to (d) large with fixed diameter $d$ and spacing $\ell$, and a cone that was either (e) inverted or (f) upright. In addition, we have plotted data for the experimentally designed surfaces described in table \ref{tab:surfaces}.}
    \label{fig:slip_and_perm_different_surfaces}
\end{figure}

To make a prediction for a given rough surface, we must compute the slip $\mathcal{L}$ and interface permeability $\mathcal{K}^{itf}$ associated to the structure of the rough features, for which we use the homogenization framework \citep{botNaq2020}. Microscopic simulations are performed as described in Appendix~\ref{appendix:microscopic_problem}. In figure~\ref{fig:slip_and_perm_different_surfaces}, we perform a parametric study of the slip and interface permeability for cylinders with varied diameter $d$ (figure~\ref{fig:slip_and_perm_different_surfaces}a, b) and varied height $h_p$ (figure~\ref{fig:slip_and_perm_different_surfaces}c, d), in addition to cones with two orientations (figure~\ref{fig:slip_and_perm_different_surfaces}e, f). We also compute $\mathcal{L}$ and $\mathcal{K}^{itf}$ for our experimental rough surfaces with parameters given in table \ref{tab:surfaces}. Figure~\ref{fig:slip_and_perm_different_surfaces}(g) shows the normalized interface permeability $\mathcal{K}^{itf}/\ell^2$ and normalized slip length $\mathcal{L}/\ell$, where $\ell$ is the size of the computational domain that defines a single periodic cell. For each structure type, we have provided an illustration of the structure with the lowest $\mathcal{L}$ and $\mathcal{K}^{itf}$ values (figure~\ref{fig:slip_and_perm_different_surfaces}a, c, e) and the structure with the highest values (figure~\ref{fig:slip_and_perm_different_surfaces}b, d, f). Except for the experimental surfaces considered (blue circle markers), the spacing $\ell$ between the structures has been kept contant. The data show an overall trend of increasing $\mathcal{K}^{itf}$ with increasing $\mathcal{L}$. Consider, for example, the case of decreasing pillar diameter $d$ (hexagonal markers): the point closest to the origin corresponds to the widest pillar (figure~\ref{fig:slip_and_perm_different_surfaces}a) and the point with highest $\mathcal{L}$ and $\mathcal{K}^{itf}$ corresponds to the thinnest pillar (figure~\ref{fig:slip_and_perm_different_surfaces}b). As the pillar diameter decreases, the slip and permeability increase, which is sensible because the amount of fluid at the equivalent surface $\mathbb{ES}$ increases as the solid fraction $\phi$ decreases.

\begin{figure}
    \centering
    \includegraphics[width=\textwidth]{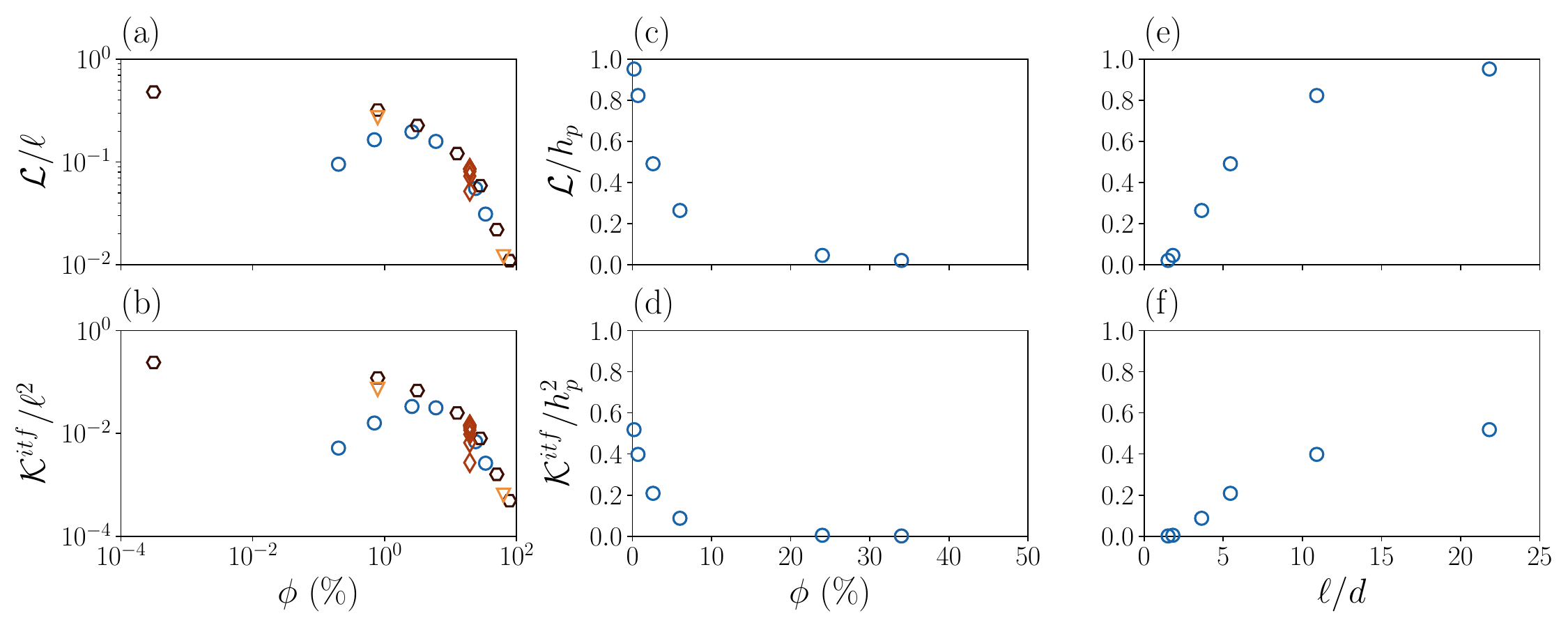}
    \caption{Nondimensional slip $\mathcal{L}/\ell$ and interface permeability $\mathcal{K}^{itf}/\ell^2$ for all the surfaces considered in figure~\ref{fig:slip_and_perm_different_surfaces} are plotted in (a, b) as a function of solid area fraction $\phi$. $\mathcal{L}$ and $\mathcal{K}^{itf}$ were normalized by $\ell$, the size of the computational domain, which is equivalent to the periodicity of the pattern. Markers are the same as in figure~\ref{fig:slip_and_perm_different_surfaces}. (c--f) Nondimensional slip $\mathcal{L}/h_p$ and interface permeability $\mathcal{K}^{itf}/h_p^2$ for surfaces with pillars described in table~\ref{tab:surfaces}, this time normalized by the pillar height $h_p$, plotted as a function of (c, d) solid area fraction $\phi$ and (e, f) normalized pillar spacing $\ell/d$. Slip and interface permeability are computed as described in Appendix~\ref{appendix:microscopic_problem}.}
    \label{fig:slip_and_perm_experimental_surfaces}
\end{figure}

Because slip and interface permeability are both derived from the structure of the rough surface \citep{boundary_conditions_naturally_permeable_wall_1967}, we expect that we cannot vary them completely as independent parameters. Indeed, early models proposed a length scale $\sqrt{k} / \alpha$ as a slip length, where $k$ is the rough/porous layer permeability and $\alpha$ is a constant depending on the particular microscopic pore structure \citep{boundary_conditions_naturally_permeable_wall_1967}. Based on this reasoning, we should then expect
\begin{equation}
    \mathcal{K}^{itf} \propto \mathcal{L}^2,
    \label{eq:K_vs_L}
\end{equation}
which we have plotted as the dash-dotted line in figure~\ref{fig:slip_and_perm_different_surfaces}. The trend with decreasing pillar diameter (hexagonal markers) seems to closely follow $\mathcal{K}^{itf} = \mathcal{L}^2$. The relationship \eqref{eq:K_vs_L} is also satisfied when considering pillars with increasing height (figure~\ref{fig:slip_and_perm_different_surfaces}, thin diamonds) and cones of two orientations (figure~\ref{fig:slip_and_perm_different_surfaces}, triangles). For comparison with previous works, we computed the value of the constant $\alpha$ for each structure and find that it varies between $\alpha \approx 0.8-2.1$, but is always of order $O(1)$ (see Appendix~\ref{appendix:alpha}), confirming the ordering relation \eqref{eq:K_vs_L}.

We also observe that the largest changes in $\mathcal{L}$ and $\mathcal{K}^{itf}$ come from changing solid fraction (figure~\ref{fig:slip_and_perm_different_surfaces}a, b or figure~\ref{fig:slip_and_perm_different_surfaces}e, f), rather than the depth of the roughness (figure~\ref{fig:slip_and_perm_different_surfaces}c, d). In figure~\ref{fig:slip_and_perm_different_surfaces}(g), large changes in $\mathcal{L}$ and $\mathcal{K}^{itf}$ are most evident with pillars of varied diameter $d$ (hexagons) and for the cones of inverted/upright orientation (triangles), compared to the case where only the pillar height $h_p$ changes (thin diamonds). Based on the trend for pillars with increasing $h_p$, it appears that the depth of a groove or structure has a much less significant effect on $\mathcal{L}$ and $\mathcal{K}^{itf}$ than changing the solid fraction $\phi$. Increasing $h_p$ has a primary effect in increasing interface permeability $\mathcal{K}^{itf}$, while the slip does not change appreciably. In physical terms, this is reasonable: the parameter $\mathcal{K}^{itf}$ describes the amount by which tangential flow at $\mathbb{ES}$ is increased by flow through the rough layer, and flow through the rough layer is less restricted when height $h_p$ grows, thus increasing $\mathcal{K}^{itf}$ until a limiting value. Slip $\mathcal{L}$ remains fairly constant because the solid fraction $\phi$ at $\mathbb{ES}$ does not change when $h_p$ changes.

Due to the increase in spacing $\ell$ between pillars, $\phi$ varies significantly for our experimental surfaces (table \ref{tab:surfaces}). To emphasize the importance of $\phi$ in determining $\mathcal{L}$ and $\mathcal{K}^{itf}$, in figures~\ref{fig:slip_and_perm_experimental_surfaces}(a,~b) we first plot values of $\mathcal{L}/\ell$ and $\mathcal{K}^{itf}/\ell^2$ for all the surfaces considered in figure~\ref{fig:slip_and_perm_different_surfaces} as a function of $\phi$. We find that all the data collapse to a single curve, with both $\mathcal{L}/\ell$ and $\mathcal{K}^{itf}/\ell^2$ decreasing with increasing $\phi$. Thus, solid fraction appears to be the dominant factor in determining $\mathcal{L}$ and $\mathcal{K}^{itf}$, since the various pillar designs appear to have similar values of $\mathcal{L}$ and $\mathcal{K}^{itf}$, as long as $\phi$ is the same. Thus, for dip coating flow, and perhaps for more general thin film flow, these data suggest that the solid fraction $\phi$ is more important to determine slip and interface permeability than the precise roughness shape. 

Because micropillars are frequently used as surface textures -- for instance, in designs for microfluidic devices -- we also separately plot $\mathcal{L}/h_p$ and $\mathcal{K}^{itf}/h_p^2$ for all the pillars listed in table~\ref{tab:surfaces} as a function of $\phi$ (figure~\ref{fig:slip_and_perm_experimental_surfaces}c,~d) and as a function of spacing normalized by the pillar diameter $\ell/d$ (figure~\ref{fig:slip_and_perm_experimental_surfaces}e,~f). Both $\mathcal{L}$ and $\mathcal{K}^{itf}$ decrease as $\phi$ increases, until they are nearly zero at $\phi \approx 30\%$. Both $\mathcal{L}/h_p$ and $\mathcal{K}^{itf}/h_p^2$ increase with increasing normalized spacing $\ell/d$. The slip length $\mathcal{L}$ appears to approach the pillar height $h_p$ ($\mathcal{L}/h_p \to 1$) at large $\ell/d$. Physically, this indicates that as $\ell/d$ becomes large, the interface at $\mathbb{ES}$ appears to be almost completely a liquid area with only a small solid fraction, and the predicted slip length appears to reflect a velocity profile over a liquid region with a no-slip condition at $x = -h_p$. The asymptotic values of slip and permeability are reached after $\ell$ is greater than approximately 10--20 pillar diameters ($\ell/d \gtrsim 10$).

\subsection{Model predictions}\label{subsec:model_predictions}

\begin{figure}
    \centering
    \includegraphics[width=\textwidth]{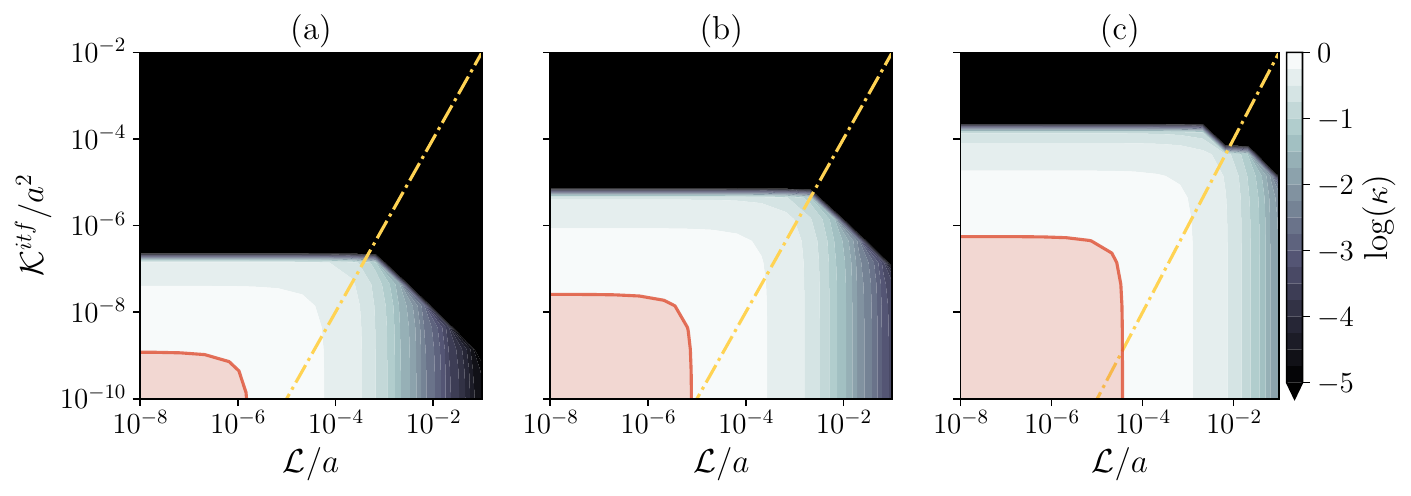}
    \caption{Curvature $\kappa$ as a function of normalized slip $\mathcal{L}/a$ and normalized interface permeability $\mathcal{K}^{itf}/a^2$, which have been rescaled with the capillary length $a$. We consider three capillary numbers: (a) $Ca = 10^{-5}$, (b) $Ca = 10^{-4}$, and (c) $Ca =  10^{-3}$. The red shaded region indicates where $\kappa \geq 0.99 \kappa_{\text{smooth}}$, where $\kappa_{\text{smooth}}$ is the curvature for the smooth case. The relation $\mathcal{K}^{itf} = \mathcal{L}^2$ is plotted (yellow dash-dotted line).}
    \label{fig:kappa}
\end{figure}

We have examined how microscopic shape of rough features affects the values of the macroscopic effective parameters $\mathcal{L}$ and $\mathcal{K}^{itf}$, which are a homogenized representation of a rough surface's properties. Now, we turn our focus toward determining how these effective parameters affect the dip coating flow over a plate (figure \ref{fig:dip_coating_system}). The solution to \eqref{eq:fixed_point_kappa} for the meniscus curvature $\kappa$ is plotted as $\log(\kappa)$ in figure~\ref{fig:kappa} for varied slip $\mathcal{L}$ and interface permeability $\mathcal{K}^{itf}$, which have been normalized by the capillary length $a$. The solution has been plotted for varied values of capillary number $Ca$, the nondimensional pulling velocity. Note that because we are now considering the macroscopic problem, we normalize the $\mathcal{L}$ and $\mathcal{K}^{itf}$ by the macroscopic scale $a$. Microscopic length scales have been `forgotten' following the homogenization procedure.

For small slip $\mathcal{L} \to 0$ or small interface permeability $\mathcal{K}^{itf} \to 0$, the film thickness given by computing \eqref{eq:free_film_thickness_solution_Ca_and_a} is found to be the same as the value expected for a smooth surface, $\kappa_{\text{smooth}} = 0.6445$ \citep{dragging_liquid_moving_plate_1942, drag_out_problem_film_coating_theory_1982}. To visualize a range of slip and permeability where the roughness may be considered negligible, we have plotted the contour given by $\kappa = 0.99 \kappa_{\text{smooth}}$ and shaded the region $0.99 \kappa_{\text{smooth}} \leq \kappa \leq \kappa_{\text{smooth}}$ in red. Within this range, curvature $\kappa$, and thus the film thickness $\bar{h}_0$, differs from that of a smooth plate by less than 1\%. The red shaded region is informative about the sensitivity of curvature $\kappa$ to varied $\mathcal{L}$ and $\mathcal{K}^{itf}$, while we note that $\kappa$ is exactly $\kappa_{\text{smooth}}$ only at the origin $(\mathcal{L}, \mathcal{K}^{itf}) = (0, 0)$. We then observe that increasing either $\mathcal{L}$ or $\mathcal{K}^{itf}$ decreases $\kappa$ and thus decreases the free film thickness $\bar{h}_0 \propto \kappa$. Another region of interest is the black region where $\kappa = 0$. By \eqref{eq:free_film_thickness_solution}, for $\kappa = 0$, we have $\bar{h}_0 = 0$ and no free film is coated, which qualitatively corresponds to the physical observation by \citet{coating_textured_solid_2011}. Thus, for a given liquid and withdrawal speed, the results in figure \ref{fig:kappa} indicate that designing a surface with a sufficiently high $\mathcal{L}$ or $\mathcal{K}^{itf}$, or an appropriate combination of the two, ensures that a film is coated only within the rough layer and not on top of the textures. From the panels with increasing $Ca$ from left to right in figure~\ref{fig:kappa}, we can make a physical observation that higher $Ca$ (having either higher speed $v_0$, higher viscosity $\eta$, or a lower liquid surface tension $\gamma$), increases the values of $\mathcal{L}$ or $\mathcal{K}^{itf}$ necessary to achieve coating only within the rough layer, without an additional free film.

\begin{figure}
    \centering
    \includegraphics[width=0.95\textwidth]{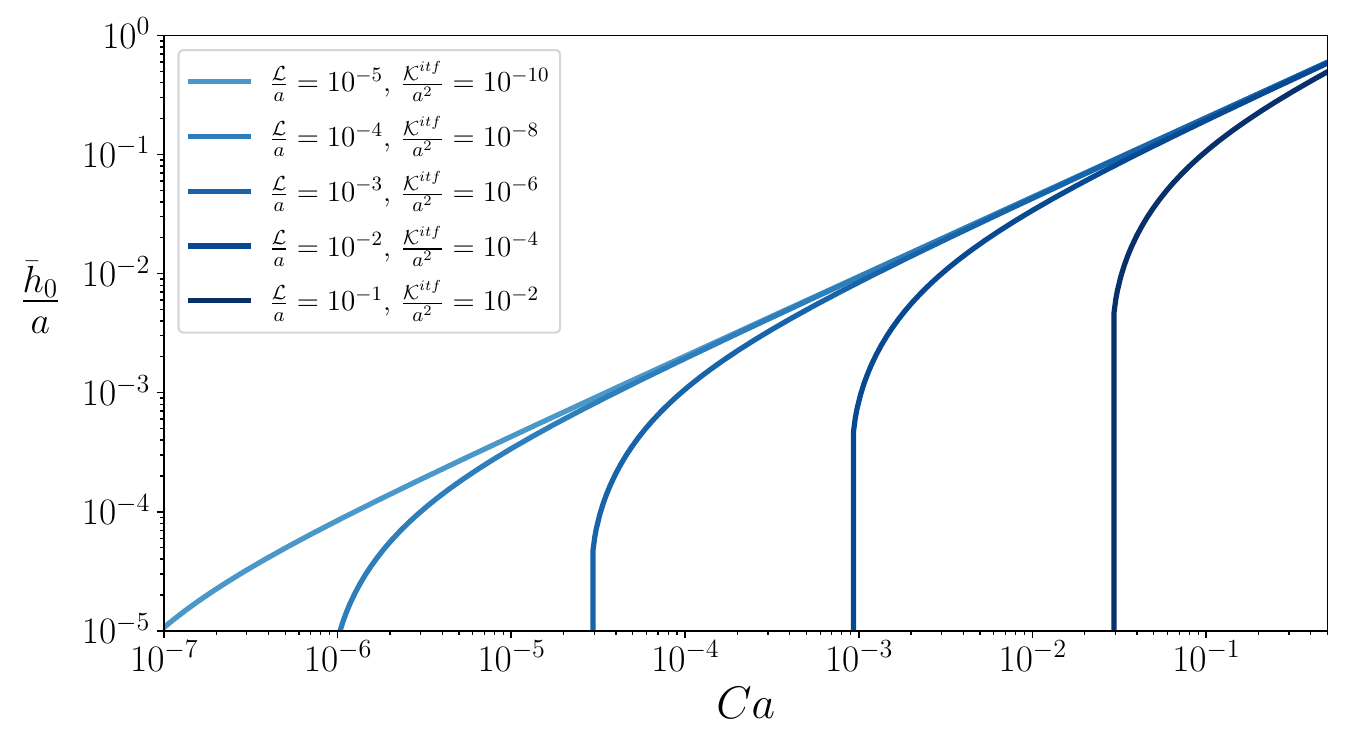}
    \caption{Dimensionless free film thickness $\bar{h}_0/a$ varies for surfaces with different normalized slip $\mathcal{L}/a$ and interface permeability $\mathcal{K}^{itf}/a^2$. Assuming that $\mathcal{K}^{itf} \propto \mathcal{L}^2$, we see that increasing the magnitude of slip $\mathcal{L}$ decreases the coated free film thickness. In addition, the critical capillary number $Ca_c$ increases significantly with greater slip. The capillary length for our experimental system is $a \approx 1.5$ mm.}
    \label{fig:h0_variation_with_slip_and_perm}
\end{figure}

For a given surface (a given point in the $\mathcal{L}$--$\mathcal{K}^{itf}$ space in figure~\ref{fig:kappa}), we can observe how the film thickness depends on the nondimensional pulling velocity $Ca$. Figure~\ref{fig:h0_variation_with_slip_and_perm} is a plot of the normalized film thickness $\bar{h}_0/a$ as a function of $Ca$, for five different surfaces with varied $\mathcal{L}$ and $\mathcal{K}^{itf}$. We have chosen values such that $\mathcal{K}^{itf} = \mathcal{L}^2$, as we have observed from the data presented in figure \ref{fig:slip_and_perm_different_surfaces}. As mentioned before, in all cases there is a critical capillary number $Ca_c$ at which no film is coated on top of the textures, in qualitative agreement with the experiments of \citet{coating_textured_solid_2011}. The $Ca_c$ value increases with increasing $\mathcal{L}$ and $\mathcal{K}^{itf}$. Physically, this is because a higher $\mathcal{L}$ or $\mathcal{K}^{itf}$ signifies less viscous stress at $\mathbb{ES}$ that would promote deposition. Therefore, it is more difficult, and requires higher velocity, to coat a free film on surfaces with roughness features promoting slip at the interface, a result that has implications for surface design.

\subsection{Comparison with experiments}\label{subsec:comparison_with_experiments}

\begin{figure}
    \centering
    \includegraphics[scale=0.50]{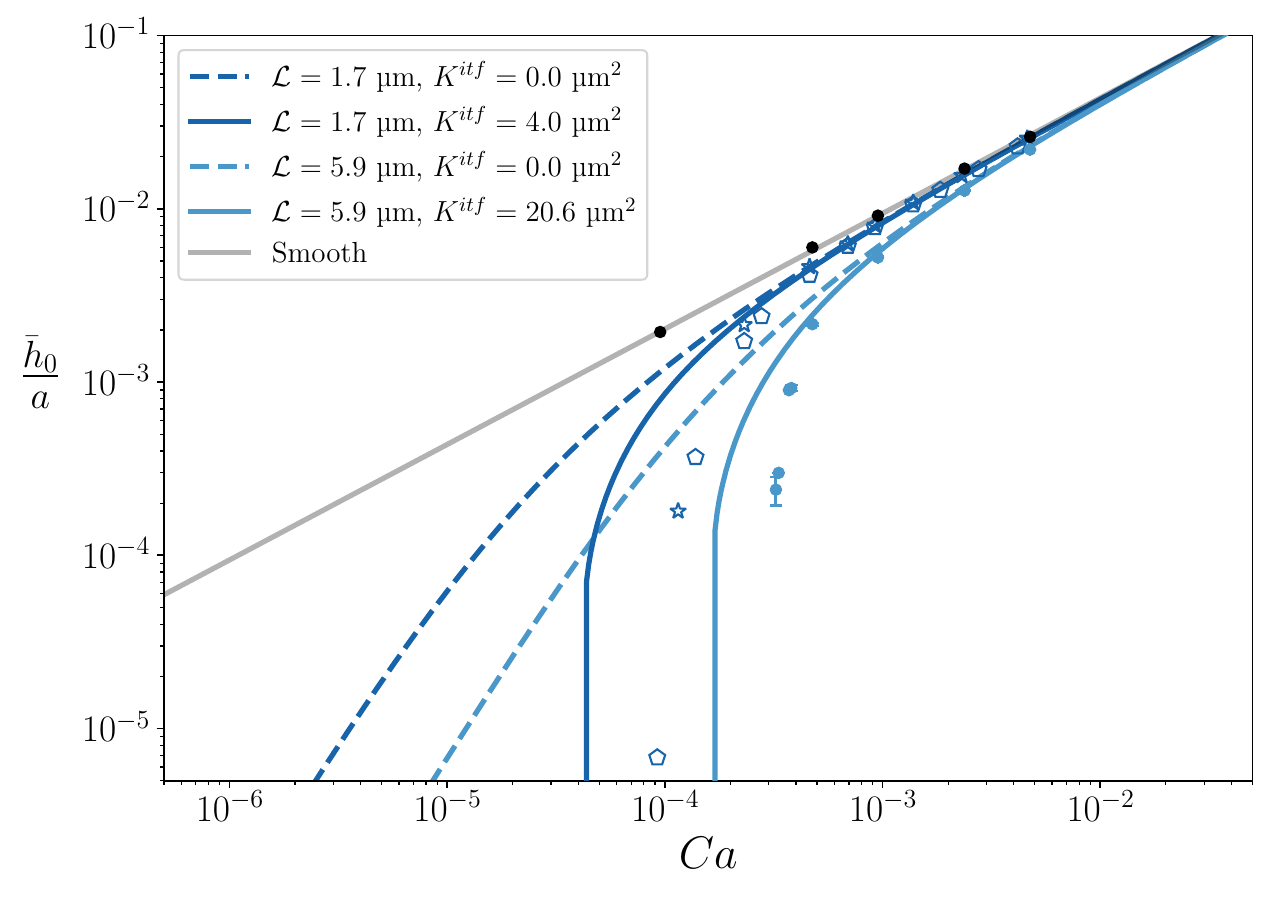}
    \caption{Measured dimensionless free film thickness $\bar{h}_0/a$ compared to model with and without $K^{itf}$. The solid gray line indicates the theoretical prediction for a smooth plate, with the solid black points indicating our experimental measurements (error bars indicate 95\% confidence intervals). Dashed blue lines indicate theoretical predictions for the `pure slip' case ($\mathcal{K}^{itf} = 0$ \textmu m$^2$), whereas solid blue lines indicate predictions with nonzero slip and interface permeability. Data points are from our experiments with a surface having pillar spacing $\ell = 36$~\textmu m (filled symbols) or from \citet{coating_textured_solid_2011} (open symbols). Shapes represent different liquid viscosities used in the experiments of \citet{coating_textured_solid_2011}: 19 mPa~s (pentagons) and 97 mPa~s (stars). The capillary length in all experiments is $a \approx 1.5$~mm.}
    \label{fig:h0_experimental}
\end{figure}

As described in Section \ref{subsec:slip_and_interface_perm_relation}, homogenization has allowed us to deduce effective parameters $\mathcal{L}$ and $\mathcal{K}^{itf}$ from arbitrary periodic microscopic structures. The microscopic simulation provides closure to the system \eqref{eq:differential_equation_xi_mu}--\eqref{eq:boundary_condition_dxi_dmu} by providing values of $\mathcal{L}$ and $\mathcal{K}^{itf}$, and in Section \ref{subsec:model_predictions} we solved the macroscopic problem for the free film thickness $\bar{h}_0$. We now use the model to predict experimentally measured film thicknesses. Predicted $\bar{h}_0$ for a surface with $\ell = 36$ \textmu m is plotted as a function of $Ca$ in figure~\ref{fig:h0_experimental} (light blue solid line), where we have $\mathcal{L} = 5.94$~\textmu m and $\mathcal{K}^{itf} = 20.61$~\textmu m$^2$. The model is in good agreement with experimental data (light blue filled symbols), although we observe a slight deviation around $Ca = 3\times 10^{-4}$ where the free film thickness sharply decreases to zero. Each experimental point represents the average of 5--6 measurements, vertical error bars represent a 95\% confidence interval in $\bar{h}_0/a$, and horizontal error bars (too small to be visible) likewise represent a 95\% confidence interval in $Ca$. For comparison, we have also plotted the data of \cite{coating_textured_solid_2011} at two viscosities (open symbols) compared to our model (dark blue solid line) where slip and interface permeability are computed for the pillar design described in \citet{coating_textured_solid_2011}, giving values $\mathcal{L} = 1.7$ \textmu m and $\mathcal{K}^{itf} = 4.0$ \textmu m$^2$. Though the prior experiments were conducted for two different viscosities (indicated by either pentagons or stars), the model prediction does not differ for varied viscosity. The model (solid lines) is able to reproduce the sharp decrease in $\bar{h}_0$ at a critical capillary number $Ca_c$, though with a deviation at low $\bar{h}_0$. The results in figure \ref{fig:h0_experimental} emphasize the importance of the interface permeability term to the present model, which arises from the form of \eqref{eq:simplified_boundary_condition_uz_dim}. Without this term (setting $K^{itf} = 0$), a model that only includes the slip term (figure~\ref{fig:h0_experimental}, dashed lines) is not able to capture the behavior at $Ca_c$. As expected, at large $Ca$, the data converge to $\bar{h}_0$ as given by \eqref{eq:smooth_surface_film_thickness}.

\begin{figure}
    \centering
    \includegraphics[width=0.8\textwidth]{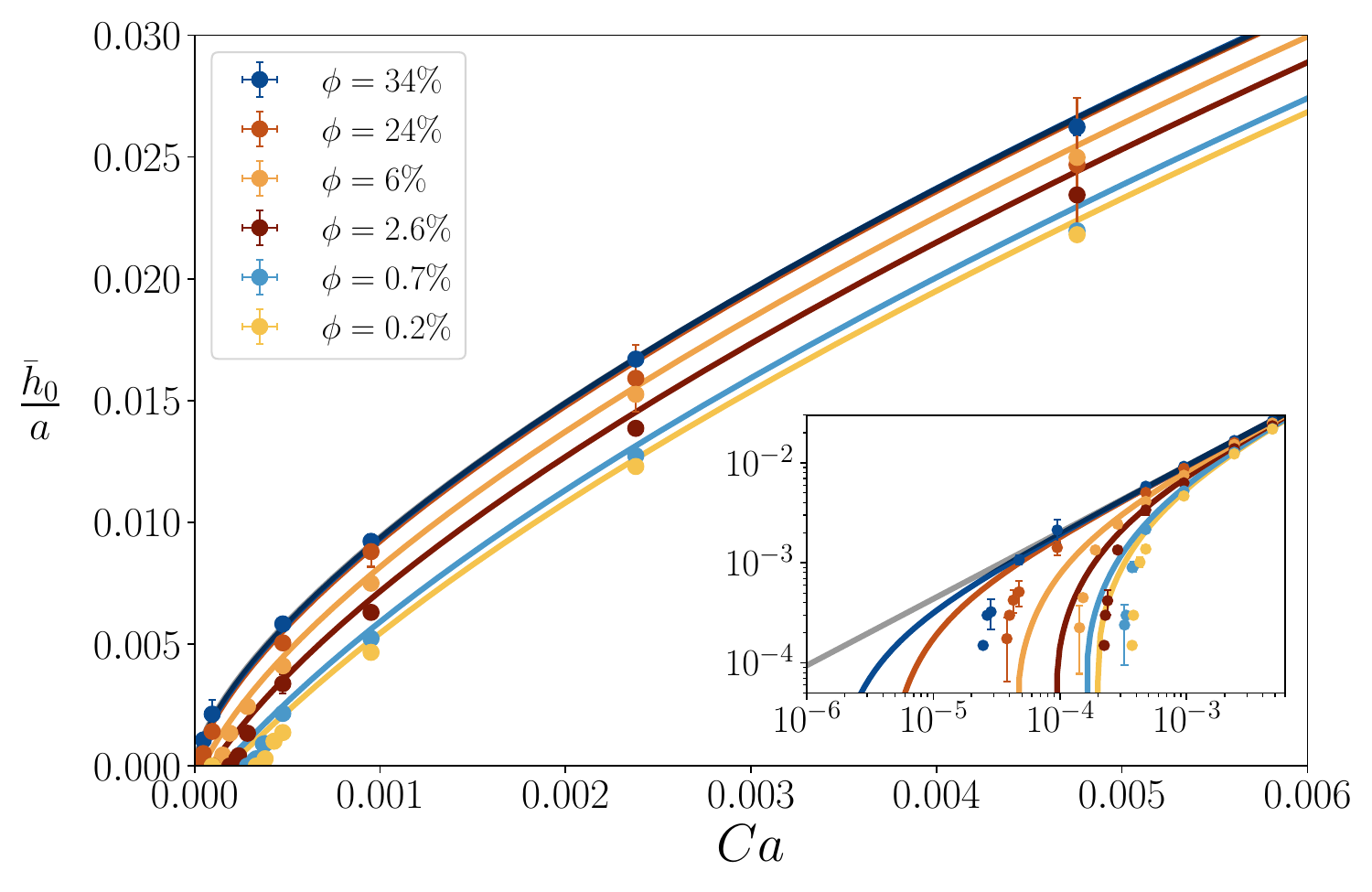}
    \caption{Dimensionless free film thickness $\bar{h}_0/a$ for surfaces with varied area fraction $\phi$. Each surface is represented by a different color, where experimental measurements are plotted as points and the model is plotted as a solid line. Error bars indicate 95\% confidence intervals. (inset) Plot on a logarithmic scale to better display data at low $Ca$.}
    \label{fig:all_surfaces_data}
\end{figure}

Our model underpredicts $Ca_c$ compared to both \citet{coating_textured_solid_2011}'s experimental data (dark blue open symbols) and our experimental data (light blue filled symbols). In the prior experiments, the critical capillary number was $Ca_c \approx 10^{-4}$ whereas the model predicts $Ca_c \approx 5 \times 10^{-5}$, so the model underpredicts the critical capillary number by about $50\%$. The solid fraction was about $\phi = 7\%$ \citep{coating_textured_solid_2011}. In our experiment $\phi = 0.7\%$, and we observe a similar deviation, with predicted $Ca_c \approx 2\times10^{-4}$ and a measured $Ca_c \approx 3\times10^{-4}$.

Coating experiments were performed for all the surfaces listed in table~\ref{tab:surfaces}. Nondimensional film thickness $\bar{h}_0/a$ is plotted as a function of $Ca$ in figure~\ref{fig:all_surfaces_data}. Each surface is indicated in a different color, where experiments are plotted as points and the model prediction is plotted as a solid line. Our model can accurately predict film thickness coated on a rough plate, and does so without fitting parameters. The inset displays the same data on a logarithmic scale to better show the data at low $Ca$ and $\bar{h}_0/a$. As may be expected, the model predicts coated film thickness accurately until the film becomes so thin that the interface approaches the top surface of the pillars; in this case, the homogenization parameter $\epsilon = \ell/\bar{h}_0$ is no longer small and the agreement with the model becomes less accurate. It is also clear from the inset of figure~\ref{fig:all_surfaces_data} that the prediction of $Ca_c$ is much better for lower $\phi$ than for dense pillar arrays. Remarkably, the model still captures the trend in the data for $\epsilon$ approaching $O(1)$. Thus, we are able to model the effect of microstructure on coating thickness in dip coating in a predictive way, and our results demonstrate a large range of validity of the homogenization approach, including its application to model thin film flows, a class of problems that may initially seem to present adverse conditions for use of the homogenization framework.

\section{Discussion and conclusion}\label{sec:discussionandconclusion}

Coating flows and wetting on rough surfaces are common in both natural and engineered systems \citep{rhomboidal_lattice_structure_common_feature_sandy_beaches_1976, purity_sacred_lotus_escape_contamination_biological_surfaces_1997, thin_film_evolution_thin_porous_layer_modeling_tear_film_contact_lens_2010}. Here we have investigated film deposition by dip coating a rough plate. One approach to modeling a rough boundary is to find equivalent averaged macroscopic properties that can be applied at a smooth equivalent surface $\mathbb{ES}$. Using this method, we show that the homogenization technique is able to predict experimentally measured film thicknesses coated on rough plates. The model overcomes the two main difficulties in modeling dip coating of rough surfaces: first, it overcomes the computational expense of a direct numerical simulation, because a macroscopic model can be used that still accurately captures the effect of microscopic structure on the flow. Second, the model is predictive, requiring no fitting parameters. In this way, the model does not require experimental data for closure and provides more insight into the physics of flow over the rough interface by employing a boundary condition that accurately models the flow. In addition, with no fitting parameters, our model is a promising method for surface design and characterization.

The model and experiments presented here provide additional physical insight into the mechanism for film thickness modification by a rough surface: both the slip and interface permeability contributions lead to less viscous stress at the boundary $\mathbb{ES}$ (see \eqref{eq:simplified_boundary_condition_uz_dim}), implying that there is less viscous force to compete with the capillary suction responsible for thinning the film. This explains the thinner free film observed than would otherwise be for a solid surface at the same plane $\mathbb{ES}$. In addition to slip $\mathcal{L}$, interface permeability $\mathcal{K}^{itf}$ contributes a flow along the interface that is driven by a tangential pressure gradient. In dip coating, the pressure gradient is a capillary pressure gradient induced by the meniscus curvature and this pressure gradient drives additional flow along the interface $\mathbb{ES}$ when $\mathcal{K}^{itf}$ is nonzero.

Previous models have been proposed to link specific texture patterns and the slip length $\mathcal{L}$, including ridges, pillars and holes \citep{effective_slip_pressure_driven_stokes_flow_2003, achieving_large_slip_superhydrophobic_surfaces_scaling_laws_generic_geometries_2007}. We have demonstrated that the microscopic structure primarily contributes to in a surface's macroscopic interfacial properties ($\mathcal{L}$, $\mathcal{K}^{itf}$) based on the structure's solid fraction $\phi$, rather than the particular shape of the rough features (see figure~\ref{fig:slip_and_perm_experimental_surfaces}). This is an important result for applications, because it implies that designing a particular surface's $\mathcal{L}$ and $\mathcal{K}^{itf}$ requires attention predominantly to the solid fraction $\phi$, whereas the particular shape of the roughness can be tuned to other design requirements.

There are many avenues for extension of the current theory, both to different coating scenarios, such as fiber coating or spin coating of rough surfaces, and to different parameter regimes. For example, corrections to model the film thickness on a smooth plate when gravity or inertia cannot be neglected may be applied to our model to extend its validity over a greater range of $Ca$ \citep{gravity_inertia_effects_plate_coating_1998}. It is also interesting to consider a qualitative comparison between our experiments and observations of `paradoxical lubrication' of drops in rough microchannels \citep{motion_viscous_droplets_rough_confinement_paradoxical_lubrication_2019}, where drops are pulled by gravity through a tilted rough microchannel. The drops are observed to descend very slowly below a critical tilt angle; above this critical angle, their descent velocity in the channel increases significantly. \citet{motion_viscous_droplets_rough_confinement_paradoxical_lubrication_2019} proposed a physical mechanism for this behavior: Below the critical velocity, almost no lubricating film separates the drop from the rough surface of micropillars. Above a critical velocity, a lubricating film separates the drop from the roughness, allowing it to move faster through the channel. These two cases correspond to experimental observations in dip coating of a rough plate, where two regimes of film thickness are observed: no free film is coated on the plate below a critical coating velocity (below $Ca_c$), corresponding to the `trapped' drop, whereas above the critical velocity (above $Ca_c$), a free film is coated on the rough plate, corresponding to the `lubricated,' quickly descending drop. In both the dip coating experiments (see figure~\ref{fig:all_surfaces_data}a) and the drop sliding experiments \citep{motion_viscous_droplets_rough_confinement_paradoxical_lubrication_2019}, the critical velocity to coat a lubricating film increases with smaller solid fraction, demonstrating an agreement in the thin film behavior in both cases. Though we here provide a qualitative comparison, we expect that there is a genuine correspondence to be explored between these systems, since dip coating and drops in microchannels are known to have a close correspondence as coating processes driven by dynamic menisci \citep{cantat2013}.

In addition, our model could be extended to describe thin film flow over many types of complex natural surfaces \citep{purity_sacred_lotus_escape_contamination_biological_surfaces_1997, characterization_distribution_water_repellent_self_cleaning_plant_surface_1997} and engineered surfaces \citep{3D_micropatterned_functional_surface_inspired_by_salvinia_molesta_2023}. For example, the model is promising for describing microfluidic mixers, in which a flow through a microfluidic channel passes over a rough pattern to promote mixing \citep{chaotic_mixer_microchannels_2002}. Currently, we have tested the model only for the case when the surface is fully wetted. However, since many types of hydrophobic surfaces are wetted in a state where air is trapped within the textures and liquid is retained only on top \citep{wettability_porous_surfaces_1944}, it would be useful to study the effective slip and interface permeability when the microscopic cell has a multiphase combination of liquid and gas \citep{effective_slip_pressure_driven_stokes_flow_2003, achieving_large_slip_superhydrophobic_surfaces_scaling_laws_generic_geometries_2007, apparent_slip_drag_reduction_flow_superhydrophobic_lubricant_impregnated_surfaces_2018}. Other works have probed the effect of anisotropic roughness on thin film flow, which is another area where the present model can be extended \citep{effective_stress_jump_across_membranes_2020}.

As we have previously mentioned, our work fits into the context of several previous works challenging the typical assumptions of homogenization, such as those that relax the assumption of periodicity \citep{effective_stress_jump_across_membranes_2020, homogenization_based_design_microstructured_membranes_2021}. Homogenization has been employed abundantly to characterize flow in large fluid domains \citep{permeability_rigid_porous_media_2010, transfer_mass_momentum_rough_porous_surfaces_2020, interfacial_conditions_free_fluid_region_porous_medium_2021}, due to the assumption requiring a large separation of scales between the macroscopic structure and the microscopic roughness scale. However, the present results suggest that a homogenized interface condition can remain predictive for thin or shallow liquid layers, demonstrating that a homogenized model is effective for thin film flow. Thus, it also seems promising to investigate applications to thin film flows in other coating or wetting scenarios, such as drop wetting on textured surfaces.

\bigskip

\noindent \textbf{Acknowledgements.} We gratefully acknowledge Dr. Sajjad Azimi for a helpful discussion regarding the transformed equation.

\noindent \textbf{Funding.} This work was supported by the Swiss National Science Foundation (G.A.Z., grant number PZ00P2\_193180).

%%%
%%% APPENDIX
%%%
\appendix

\section{Transformation from $\mu(\lambda)$ to $\xi(\mu)$}\label{appendix:transformation}

To understand the nature of the transformation \eqref{eq:lambda_to_xi_transformation}, note that \eqref{eq:lambda_to_xi_transformation} gives
\begin{equation}
    \frac{d\mu}{d\lambda} = -\sqrt{\xi},
\end{equation}
where we must take the negative root in $\pm \sqrt{\xi}$ because we know that the slope of the film profile is always negative (see figure~\ref{fig:dip_coating_system}b). The higher derivatives are then derived as
\begin{align}
    \frac{d^2\mu}{d\lambda^2} &= \frac{d}{d\lambda} \left(-\sqrt{\xi}\right) = \frac{d\mu}{d\lambda} \frac{d}{d\mu} \left(-\sqrt{\xi}\right) = \frac{1}{2} \frac{d\xi}{d\mu},
    \label{eq:derivation_of_d2mu_dlambda2} \\
    \frac{d^3\mu}{d\lambda^2} &= \frac{d}{d\lambda} \left(\frac{d^2\mu}{d\lambda^2}\right) = \frac{d\mu}{d\lambda} \frac{d}{d\mu} \left(\frac{1}{2}\frac{d\xi}{d\mu}\right) = -\frac{1}{2} \frac{d^2\xi}{d\mu^2} \sqrt{\xi}.
    \label{eq:derivation_of_d3mu_dlambda3}
\end{align}
Equation \eqref{eq:derivation_of_d3mu_dlambda3} is the basis of the transformed equation \eqref{eq:differential_equation_xi_mu}.

\section{Calculation of slip and interface permeability}\label{appendix:microscopic_problem}

\begin{figure}
    \centering
    \includegraphics[width=\textwidth]{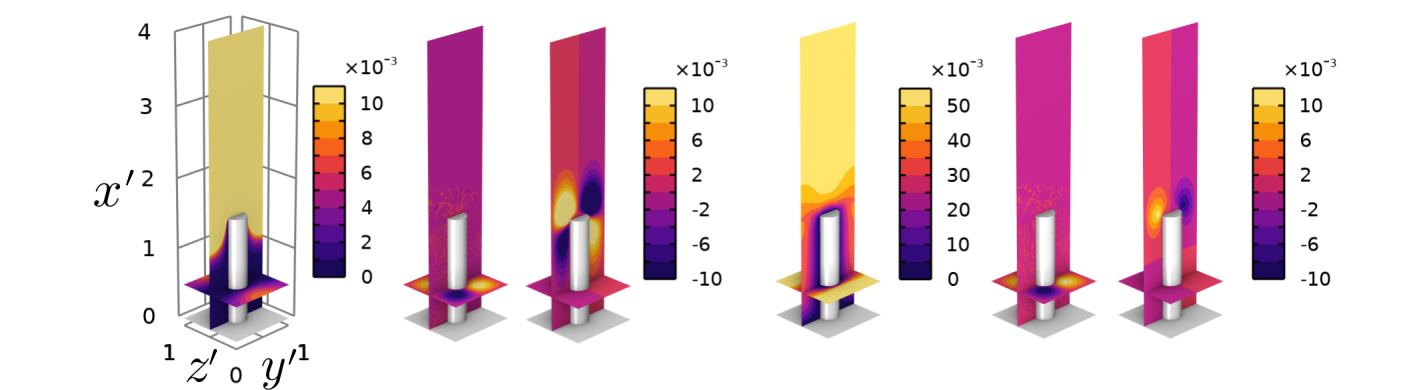}
    \caption{Vector components of $\boldsymbol{\lambda}$ and $\boldsymbol{\psi}$ within the microscopic domain. From left to right the colours represent the isocontours of $\lambda_{z'}$, $\lambda_{y'}$, $\lambda_{x'}$, $\psi_{z'}$, $\psi_{y'}$, and $\psi_{x'}$ on the planes $\{z'=0.5\ell\}$ and $\{x'=0.5\ell\}$.}
    \label{fig:microscopic_problem}
\end{figure}

Several works based on the multiscale analysis of flows at the interface between a free-fluid region and a micro-structured substrate have been developed in recent years \citep[see, for instance,][]{jimbol2017,generalized_slip_condition_over_rough_surfaces_2019, transfer_mass_momentum_rough_porous_surfaces_2020}. \cite{botNaq2020,interfacial_conditions_free_fluid_region_porous_medium_2021} wrote the approximation of the macroscopic velocity field up to order $\epsilon^3$. The estimation of the macroscopic velocity in the $z$-direction (equation \ref{eq:simplified_boundary_condition_uz_dim}) depends on the slip and interface permeability coefficients, $\mathcal{L}$ and $\mathcal{K}^{itf}$, which are calculated from the solution of closure problems valid in the microscopic elementary cell shown in figure~\ref{fig:microscopic_problem}:
\begin{equation}\label{a1}
\nabla\cdot\boldsymbol{\lambda}=0, \quad-\nabla \xi+\nabla^2\boldsymbol{\lambda}=0, \\
\end{equation}
\begin{equation*}
\lim _{x'\rightarrow+\infty} \frac{\partial \lambda_{z'}}{\partial x'}=1, \quad \lim _{x' \rightarrow+\infty} \xi=0, 
\end{equation*}
and 
\begin{equation}\label{a2}
\nabla\cdot\boldsymbol{\psi}=0, \quad-\nabla \chi+\nabla^2\boldsymbol{\psi}=H^*(-x'+h_p),
\end{equation}
\begin{equation*}
\begin{split}
\lim _{x' \rightarrow+\infty} \frac{\partial \psi_{z'}}{\partial x'}=0, \quad \lim _{x' \rightarrow+\infty} \chi=0.
\end{split}
\end{equation*}
where the triplet $(x',y',z')$ represents the microscopic spatial variables related to $(x,y,z)$ by $(x',y',z')=(x,y,z)/\epsilon$ \citep{botNaq2020}.
The variables $\boldsymbol{\lambda}$, $\boldsymbol{\psi}$, $\xi$ and $\chi$ are auxiliary microscopically periodic variables introduced during the homogenization procedure, 
while $H^*$ is the Heaviside function centered in $x' = h_p$, corresponding to the tip of the protrusions forming the structured substrate. The macroscopic quantities $\mathcal{L}$ and $\mathcal{K}^{itf}$ used in the interface conditions \eqref{eq:simplified_boundary_condition_uz_dim} are then retrieved by the solutions of problems \eqref{a1}--\eqref{a2} introducing the following relations
\begin{equation}\label{avg1}
   \mathcal{L}= - x' + h_p + \int_0^1\lambda_{z}(x'\rightarrow+\infty)\,dz \quad \text{and} \quad
   \mathcal{K}^{itf}=\int_0^1\psi_{z}(x'\rightarrow+\infty)\,dz.
\end{equation}
Problems \eqref{a1}--\eqref{a2} have been solved numerically for the substrate geometries shown in figure~\ref{fig:surfaces}. The numerical solution relies on a weak form implementation in the finite-element solver COMSOL Multiphysics. The spatial discretization is based on P1-P2 Taylor-Hoods elements for the couples $(\xi,\boldsymbol{\lambda})$ and $(\chi,\boldsymbol{\psi})$. We use mesh spacing $\Delta l_1=0.1$ at the boundaries of the microscopic cell and we guarantee at least $10$ grid points on each side of the solid inclusions when the spacing $\Delta l_1$ produces less than 10 points on that side. Other simulations have been carried out on finer meshes with spacing $\Delta l_2=\Delta l_1/2$ and $\Delta l_3=\Delta l_1/4$ and numerical convergence of the average values of $\mathcal{L}$ and $\mathcal{K}^{itf}$ up to $2\%$ has been verified between $\Delta l_2$ and $\Delta l_3$.
As a matter of example, figure~\ref{fig:microscopic_problem} shows the microscopic fields associated with the structure used to calculate the light blue solid profile of figure~\ref{fig:h0_experimental}. 

\section{Relation between $\mathcal{L}$ and $\mathcal{K}^{itf}$}\label{appendix:alpha}

As discussed in Section~\ref{sec:results}, early models proposing a slip length at a rough interface related its value to the permeability by a constant $\alpha$, depending on the pore structure \citep{boundary_conditions_naturally_permeable_wall_1967}. Based on the data of figure~\ref{fig:slip_and_perm_different_surfaces}, we can compute $\alpha$ for various surface designs. The value of $\alpha$ is observed to be approximately $\alpha = 1$, except when $\phi$ becomes large, where it appears that $\sqrt{\mathcal{K}^{itf}}$ grows faster than $\mathcal{L}$.

\begin{figure}
    \centering
    \includegraphics[width=0.65\textwidth]{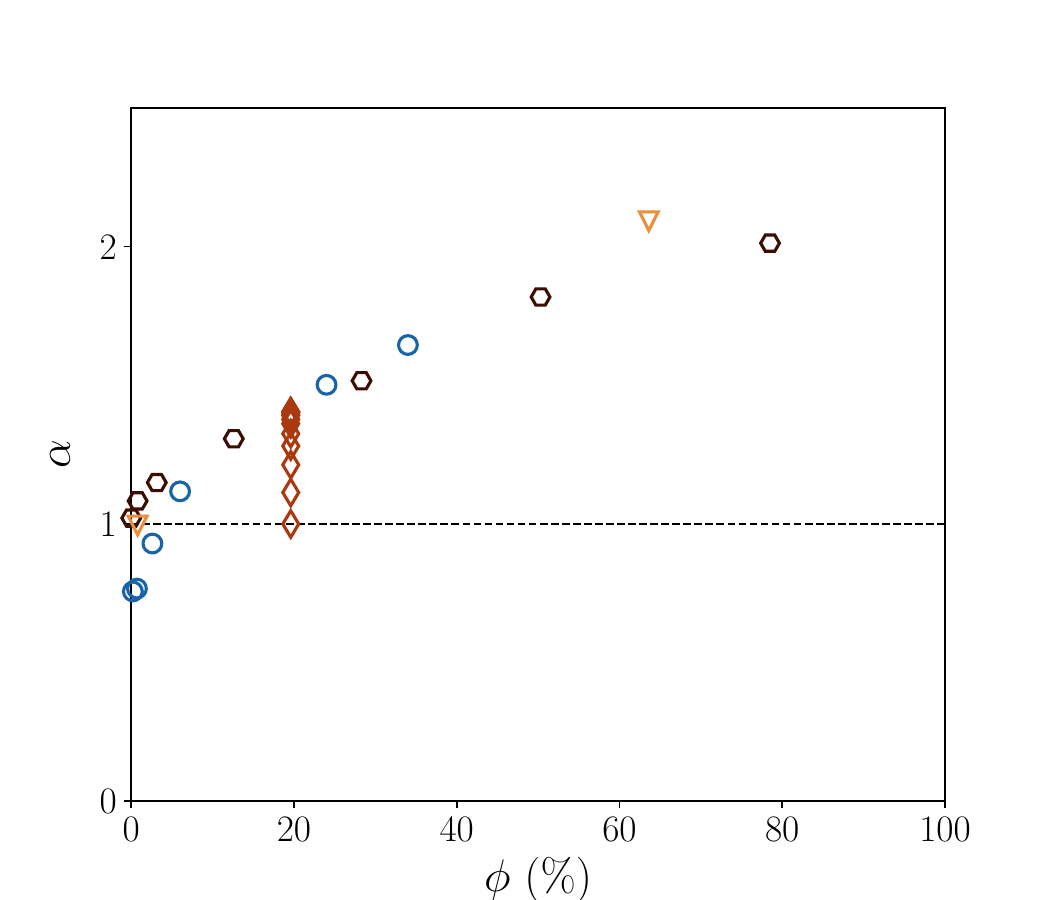}
    \caption{Constant $\alpha$ relating slip to interface permeability by $\alpha = \sqrt{\mathcal{K}^{itf}} / \mathcal{L}$ \citep{boundary_conditions_naturally_permeable_wall_1967}. Markers are the same as in figure~\ref{fig:slip_and_perm_different_surfaces}.}
    \label{fig:alpha}
\end{figure}

\FloatBarrier

\bibliography{main} 
\bibliographystyle{apalike}

\end{document}